\documentclass[journal]{IEEEtran}
\IEEEoverridecommandlockouts

\usepackage{cite}
\usepackage{amsmath,amssymb,amsfonts,bm}
\usepackage{algorithmic}
\usepackage{algorithm}
\usepackage{graphicx}
\usepackage{textcomp}
\usepackage[table]{xcolor}
\usepackage{xcolor}
\usepackage{wrapfig}
\usepackage{multirow}
\usepackage{tabularx}
\usepackage{url}
\usepackage{import}    
\usepackage{comment}  
\usepackage{dirtytalk}  
\usepackage{stfloats}
\usepackage{soul}
\usepackage{adjustbox}
\usepackage{xfrac}
\usepackage{makecell}
\usepackage{epsfig}
\usepackage{epsf}
\usepackage[hidelinks]{hyperref}
\usepackage{tcolorbox}
\usepackage{mdframed}
\usepackage{subcaption} 

\definecolor{mygray}{RGB}{240,240,240} 
\definecolor{lblue}{RGB}{237, 247, 255} 
\definecolor{lred}{RGB}{248, 232, 232} 

\def\BibTeX{{\rm B\kern-.05em{\sc i\kern-.025em b}\kern-.08em
    T\kern-.1667em\lower.7ex\hbox{E}\kern-.125emX}}

\usepackage{etoolbox}
\usepackage{tikz}

\newrobustcmd*{\fillsquare}[1]{\tikz{\filldraw[draw=#1,fill=#1] (0,0)
rectangle (0.25cm,0.25cm);}}

\newrobustcmd*{\mycircle}[1]{\tikz{\filldraw[draw=#1,fill=#1] (0,0) circle [radius=0.1cm];}}

\newrobustcmd*{\mytriangle}[1]{\tikz{\filldraw[draw=#1,fill=#1] (0.3cm,0.3cm) -- (0cm,0.3cm) -- (0.15cm,0cm);}}

\newrobustcmd*{\fillnabla}[1]{\tikz{\filldraw[draw=#1,fill=#1] (0.2cm,0) -- (0.1cm,0.2cm) -- (0,0);}}

\def\BibTeX{{\rm B\kern-.05em{\sc i\kern-.025em b}\kern-.08em
    T\kern-.1667em\lower.7ex\hbox{E}\kern-.125emX}}
\begin{document}

\title{Channel-Aware Holographic Decision Fusion}

\author{Domenico Ciuonzo,~\IEEEmembership{Senior Member,~IEEE}, Alessio Zappone,~\IEEEmembership{Fellow,~IEEE} and Marco Di Renzo,~\IEEEmembership{Fellow,~IEEE}

\thanks{Manuscript received 30th December 2024; revised 22th March 2025; accepted 1st May 2025.}
\thanks{D. Ciuonzo is with the Department of Electrical Engineering and Information Technologies (DIETI) at University of Naples Federico II, Italy (domenico.ciuonzo@unina.it).}
\thanks{A. Zappone is with Department of Electrical and Information Engineering ``Maurizio Scarano'', University of Cassino and Southern Lazio, Italy and with Consorzio Nazionale Interuniversitario per le Telecomunicazioni (CNIT), Italy (alessio.zappone@unicas.it).}
\thanks{M. Di Renzo is with Universit\'e Paris-Saclay, CNRS, CentraleSup\'elec, Laboratoire des Signaux et Syst\`emes, 3 Rue Joliot-Curie, 91192 Gif-sur-Yvette, France. (marco.di-renzo@universite-paris-saclay.fr), and with King's College London, Centre for Telecommunications Research -- Department of Engineering, WC2R 2LS London, United Kingdom (marco.di\_renzo@kcl.ac.uk).}
}

\markboth{IEEE Internet of Things Journal,~Vol.~*, No.~*, Month~2024}{Ciuonzo \MakeLowercase{\textit{et al.}}: Channel-Aware Holographic Decision Fusion}

\maketitle
\begin{abstract}
This work investigates Distributed Detection (DD) in Wireless Sensor Networks (WSNs) utilizing channel-aware binary-decision fusion over a shared flat-fading channel. 
A reconfigurable metasurface, positioned in the near-field of a limited number of receive antennas, is integrated to enable a holographic Decision Fusion (DF) system. This approach minimizes the need for multiple RF chains while leveraging the benefits of a large array.
The optimal fusion rule for a fixed metasurface configuration is derived, alongside two suboptimal joint fusion rule and metasurface design strategies. These suboptimal approaches strike a balance between reduced complexity and lower system knowledge requirements, making them practical alternatives.
The design objective focuses on effectively conveying the information regarding the phenomenon of interest to the FC while promoting energy-efficient data analytics aligned with the Internet of Things (IoT) paradigm.
Simulation results underscore the viability of holographic DF, demonstrating its advantages even with suboptimal designs and highlighting the significant energy-efficiency gains achieved by the proposed system.
\end{abstract}


\section{Introduction}
\label{Introduction}
\subsection{Context and Motivation}

\IEEEPARstart{T}{he} Internet of Things (IoT) is shaping a future where a vast network of compact devices with sensing, processing, and communication capabilities seamlessly integrates into everyday life~\cite{Ciuonzo2019book}.
Wireless Sensor Networks (WSNs) are fundamental to this vision, serving as the ``sensing arm'' of IoT. 
At the heart of WSN operations lies Distributed Detection (DD)-—a key enabler for accurate decision-making in decentralized systems.
DD within WSNs supports several vertical applications ranging from cognitive radio~\cite{Chen2019} to industrial automation~\cite{tabella2024bayesian}.

To achieve effective DD, data from dispersed sensors must be collected and transmitted efficiently. This process hinges on optimizing wireless channel usage, where factors like the broadcast nature of wireless signals, sensor observation correlations, and the demand for spectral efficiency come into play. 
These factors have prompted the shift toward DD over Multiple-Access Channels (MACs), where multiple sensors can \emph{simultaneously transmit} over a shared channel~\cite{li2007distributed}. 
However, concurrent transmissions introduce nontrivial interference beyond conventional fading effects.
Distributed Multiple-Input Multiple-Output (MIMO) techniques address these challenges, enabling \emph{receive diversity} and improving robustness to fading~\cite{zhang2008optimal,Ciuonzo2012}.
By equipping the Fusion Center (FC) with multiple antennas, the system effectively realizes a \say{virtual} MIMO channel—akin to the uplink in multiuser MIMO systems~\cite{jiang2007multiuser}.
This \say{MIMO-enabled} fusion framework improves signal reliability and ultimately strengthens collective WSN performance.

The advent of \emph{massive MIMO} technology~\cite{lu2014overview} revolutionized wireless systems by leveraging large antenna arrays to deliver notable gains in capacity, energy efficiency, and interference suppression.
Its mathematical advantages, such as small-scale fading averaging and channel orthogonalization, made it ideal for energy-constrained WSNs~\cite{ciuonzo2015,jiang2015massive}.
However, fully-digital or hybrid implementations~\cite{Chawla2021} impose high computational, hardware, and energy demands due to the need for RF chains per antenna. Additionally, practical constraints like half-wavelength spacing and mutual coupling limit the array size at the FC and thus fusion performance. 
These challenges have driven interest toward alternative, \emph{cost-effective architectures that retain the core benefits of massive MIMO while mitigating its practical overhead}.

A promising alternative lies in compact \emph{metasurfaces}~\cite{direnzo2020} composed of sub-wavelength passive and reconfigurable reflective elements which enable superior beam-steering, enhancing wireless communication coverage, spectral efficiency, and energy consumption.
This kind of solution offers versatile and energy-efficient designs and can be categorized into \emph{two scenarios}~\cite{huang2020holographic}:
($i$) metasurfaces integrated as part of the wireless channel (\emph{Reconfigurable Intelligent Surfaces}, RISs) and ($ii$) metasurfaces embedded within transceiver architectures (\emph{Reconfigurable Holographic Surfaces}, RHSs).
In the first scenario, RISs serve as intelligent reflectors, redirecting signals toward distant receivers.
This makes them well-suited for enhancing coverage, mitigating interference, and enabling non-line-of-sight communication—key applications where reshaping the wireless environment is advantageous. 
RIS-based beam steering depends on phase tuning across nearly-passive elements, typically orchestrated by an external controller.

In the second scenario, RHSs are integrated directly into transceiver systems, enabling ultra-thin, energy-efficient, and low-complexity designs~\cite{jamali2020intelligent,interdonato2024approaching} that demand precise control over the transmitted signal. 
The holographic design naturally facilitates beamforming by leveraging wave interference patterns, achieving desired radiation characteristics with minimal external control.

\begin{figure}[ht]
\includegraphics[width=1\columnwidth]{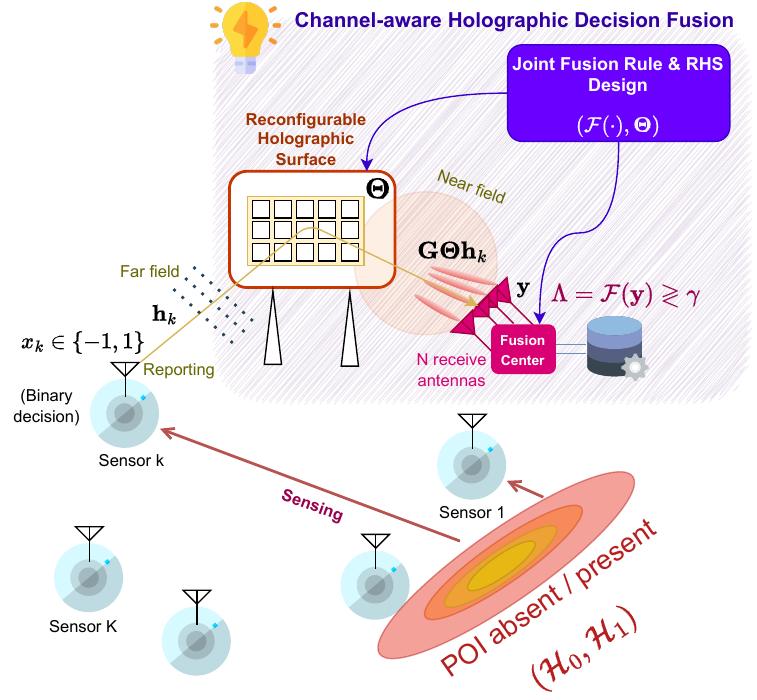}\caption{The (channel-aware) holographic DF system model considered in this work.}\label{fig.system.model}
\end{figure}

This work centers on the second category: RHS-aided receive antenna architectures designed to support DD and their corresponding system optimization.
The goal is to determine whether these architectures—through the joint optimization of RHS phase shifts and the fusion rule at FC—can achieve comparable performance as fully-digital massive MIMO-based DD systems~\cite{ciuonzo2015,dey2020wideband}.
The proposed approach, here referred to as \emph{(channel-aware) holographic DF} (illustrated in Fig.~\ref{fig.system.model}), is designed to leverage cost-effective hardware while exploiting near-field DoF for superior fusion performance. 
Unlike far-field MIMO, where spatial multiplexing is plane-wave-limited, near-field communication harnesses wavefront curvature and spherical modes for enhanced spatial resolution~\cite{dardari2020communicating}.
The RHS acts as an analog pre-processor, performing wavefront shaping and ultra-dense spatial filtering.
Sparse feeds serve as electromagnetic sensors, reducing digital beamforming complexity. 
In the context of DF, the increased DoFs translates to a larger set of equivalent channels, enabling to approach optimal (channel-free) DD.

This marks a departure from \emph{conventional (channel-aware) DF}, which employs a fully-digital architecture where each antenna in the receive array is individually connected to an RF chain~\cite{Ciuonzo2012,ciuonzo2015}, resulting in significantly higher hardware costs and energy consumption.

\subsection{Related Works}
\noindent
\textbf{Background on DD via WSNs:} the literature on DD can be categorized into three ``waves''. 
The \emph{first wave} traces back to the foundational work of Tenney and Sandell~\cite{Tenney1981}, with key contributions from~\cite{chair1986,reibman1987}.
Early studies focused on single- or multi-bit quantization of measurements/likelihoods and the design of local detectors or fusion rules, typically assuming decoupled or noise-free reporting channels.

The \emph{second wave} emerged in the early 2000s with the rise of WSNs. 
Research shifted to channel-aware fusion rules~\cite{Chen2004} and diverse reporting protocols, such as type-, time-~, frequency-, and code-based approaches~\cite{Mergen2007,Yiu2008}. 
Efforts to boost performance also introduced power allocation~\cite{zhang2008}, censoring schemes~\cite{Appadwedula2005}, and sensor subset selection~\cite{ahmadi2009}.
Early WSN setups utilized parallel-access channels, where sensors transmitted binary decisions to a FC over non-interfering channels. This setup established a foundation for fusion rule design but faced limitations in spectral efficiency and collection time.
The shift to MACs introduced the \say{virtual} MIMO system concept, leveraging multiple antennas at the FC to counteract fading while maintaining sensor simplicity~\cite{zhang2008optimal,Ciuonzo2012}.
However, concurrent sensor transmissions led to interference, necessitating efficient fusion rule designs.

In recent years, the \emph{third wave} of DD has attempted to capitalize novel concepts, such as the use of green, near-zero-energy sensors (e.g. exploiting backscattering~\cite{Ciuonzo2019pimrc} and energy-harvesting~\cite{Tarighati2017}),  millimeter Wave (mmWave) sensors~\cite{Chawla2021} and massive MIMO~\cite{ciuonzo2015,chawla2019} to \emph{reduce the energy expenditure of the WSN} by achieving desired performance.

\noindent
\textbf{Related studies on RIS-assisted WSNs:} Recent advancements in WSNs have \emph{solely} explored the potential of RISs to create \say{smarter} channel environments~\cite{zappone2022surface}.
Recent efforts in this direction included contributions in \emph{over-the-air computation (AirComp)} and \emph{inference-oriented scenarios}.

AirComp refers to the computation of \say{nomographic} functions, with the average being a classic example, by exploiting the signal superposition property of wireless multiple access channels~\cite{Sahin23}.
In this domain, RIS design—alongside other sensor and FC parameters—has been optimized to minimize the Mean Squared Error (MSE) between the computed signal and the desired nomographic function~\cite{fang2021,zhang2022worst,zhai2022beamforming,zhao2023ris}.
For instance, \cite{fang2021} introduced a framework for designing RIS phase shifts and beamforming vectors using alternating optimization and successive convex approximation.
Subsequently, \cite{zhang2022worst} addressed an RIS-assisted AirComp system under additive bounded uncertainty in channel state information (CSI), focusing on worst-case MSE under a total power constraint.
Earlier works on RIS-assisted AirComp systems typically depend on real-time CSI, which leads to excessive overhead, especially when the number of RIS elements is large. 
To address this challenge,~\cite{zhai2022beamforming} introduced a two-timescale penalty dual-decomposition algorithm.
This approach optimizes short-term transmit power and receive beamforming using low-dimensional, real-time CSI, while updating the RIS beamforming matrix based on long-term channel statistics. 
Similarly,~\cite{zhao2023ris} presented a CSI transmission-free data fusion and timing resynchronization approach for asynchronous RIS-enhanced AirComp systems.

Advanced AirComp systems have expanded to include \emph{active} RISs (accounting for self-interference effects)~\cite{zhang2023beamforming}, \emph{multi-RIS setups}~\cite{li2021double,zhai2022joint} and \emph{Simultaneously Transmitting \& Reflecting RISs} (STAR-RISs)~\cite{zhai2023simultaneously}.
The use of AirComp is also explored with mmWave communications~\cite{hu2022ris}, where RISs address channel condition limitations and blockage issues.
An optimization problem is formulated through joint design of receive beamforming, device transmit scalars, and RIS phase shifts and solved via the Riemannian conjugate gradient algorithm.
The RIS optimization framework has also been explored to leverage \emph{external energy sources}, in an ($i$) use-while-store or ($ii$) store-then-use fashion.
In the former case, RISs have been used with backscatter-based WSNs~\cite{mao2022intelligent}, with joint optimization of power splitting ratios, normalization factors, and RIS phase shifts.
In the latter case, RISs have been employed to optimize downlink energy beamforming and AirComp~\cite{wang2021wireless} or hybrid AirComp/communication tasks~\cite{mao2024joint}.
It is worth noting that AirComp-focused optimization primarily assumes ($a$) predefined computation functions and ($b$) relies on measurements that are independent, zero-mean, and unit-variance.

While much of the focus has been on enabling (nomographic) aggregation functions via RISs, relatively few works delve into \emph{inference-oriented RIS design}~\cite{Ahmed2022,ge2024ris,rajput2024joint,mudkey2022wireless,Ciuonzo2025icassp}.
For example, \cite{Ahmed2022} proposed a joint transmit and passive RIS beamforming strategy for secure parameter estimation in the presence of eavesdroppers, with a single-antenna FC. 
Similarly, \cite{rajput2024joint} explored the joint design of sensor precoders and RIS reflection matrices to minimize MSE in vector parameter estimation using alternating optimization techniques. 
In~\cite{ge2024ris}, multiple RISs are employed to optimize cooperative spectrum sensing (based on data or decision fusion logics) by enhancing the sensing phase through RIS design, using either instantaneous or statistical CSI. 

\noindent
\textbf{Related studies on RIS-assisted DD:} When it comes to a \emph{DD task} aided by RIS, the closest works are~\cite{mudkey2022wireless} and~\cite{Ciuonzo2025icassp}. 
The former presents a joint design of fusion rule and RIS shifts, assuming instantaneous CSI and ideal sensor conditions~\cite{mudkey2022wireless}. 
The latter proposes a two-timescale optimization approach for scenarios involving large arrays at the FC~\cite{Ciuonzo2025icassp}.
However, both approaches focus exclusively on RIS deployment and its optimization for the link between the sensors and FC, whether in AirComp or inference-oriented (viz. goal-oriented) setups.
The potential of metasurfaces, specifically in the form of RHS, as an enabler of holographic transceiver designs remains unexplored in DD contexts.

\noindent
\textbf{Literature gap and positioning} -- Despite the significant progress in employing RISs within WSNs, \emph{no studies to date have explored inference problems, including DD, that incorporate RHS as an integral component of the FC architecture in a holographic configuration, particularly with a goal-oriented focus and under complexity constraints}. Furthermore, while considerable efforts have been dedicated to advancing holographic transceivers for wireless communication, their potential applicability and feasibility in inference-driven WSN setups remain unaddressed, to the best of our knowledge.

\subsection{Summary of Contributions}

Accordingly, the \emph{main contributions} of this work are outlined as follows:
\begin{itemize}
\item We study DD with sensors transmitting their decisions to a FC over a multiple-access channel.
Departing from traditional fully-digital MIMO FCs that require an RF chain for each antenna, we position an RHS near the FC. This arrangement establishes a holographic multi-antenna system operating in the near-field, slashing hardware complexity and costs. Our proposed system targets high WSN detection performance with minimal complexity, cost, and power consumption.
\item To overcome the infeasibility of the Log-likelihood Ratio (LLR) and the lack of theoretical performance metrics for RHS design, we introduce \emph{two design strategies for joint fusion rule \& RHS design}.
Both approaches couple the fusion rule and RHS configuration by enforcing the fusion rule to be a widely-linear vector. 
The first strategy leverages the complete statistical characterization of the received signal vector. The second approach, based on the \say{Ideal Sensors} (IS) assumption~\cite{Chen2004,Ciuonzo2012,ciuonzo2015}, follows a pragmatic rationale, delivering an efficient joint design while remaining agnostic to sensor performance.

\item We address the two associated optimization problems via Alternating Optimization (AO) and Majorization-Minimization (MM)~\cite{sun2016majorization}. The derived solutions are also analyzed in terms of computational complexity and knowledge demands, providing practical insights into the feasibility of the proposed designs.

\item We validate the effectiveness of our holographic DF system through simulations, comparing its performance against a fully-digital massive MIMO-based DD counterpart. The proposed system achieves comparable detection performance while maintaining the benefits of reduced complexity, cost, and power consumption.
\end{itemize}

\noindent
\textbf{Practical advantages \& applications of Holographic DF:} the proposed architecture is inherently versatile, supporting a wide range of DD applications, including spectrum hole detection~\cite{salvorossi2013} and oil spill monitoring in the oil \& gas sector~\cite{tabella2024bayesian}.
Beyond its broad applicability, the holographic DF system can achieve high-accuracy inference while ensuring timely/bandwidth-efficient DD and controlled (sensor) energy usage, aligning with the green IoT vision~\cite{albreem2017green}.
Just as important, the cost-effectiveness of the holographic FC makes it an appealing choice for intermediate collector nodes, such as those in clustered WSN architectures~\cite{aldalahmeh2022}.

\subsection{Paper Organization and Notation}

The rest of the paper is organized as follows.
Sec.~\ref{Model} introduces the system model. Sec.~\ref{sec_fusion_rule_RIS} formulates the joint fusion \& RHS design problem (supporting the proposed holographic DF system), while Sec.~\ref{sec:design_solution} describes our AO-based solution.
Our approach is then evaluated via simulations in Sec.~\ref{Sim_res}.
Sec.~\ref{Conclusions} ends the paper with some pointers to research prospects.\\
\textbf{Notation} -- vectors (resp. matrices) are denoted with lower-case (resp. upper-case) bold letters; $\mathbb{E}\{\cdot\}$,
$\mathrm{var\{\cdot\}}$, $\mathrm{Cov(\cdot)}$, $\mathrm{PCov(\cdot)}$, $(\cdot)^{T}$, $(\cdot)^{\dagger}$, $\Re\left(\cdot\right)$, $\angle(\cdot)$ and $\left\Vert \cdot\right\Vert $ denote expectation,
variance, covariance, pseudocovariance, transpose, conjugate transpose, real part, phase, and Euclidean norm operators, respectively; $\bm{O}_{N\times K}$
(resp. $\bm{I}_{N}$) denotes the $N\times K$ (resp. $N\times N$)
null (resp. identity) matrix; $\bm{0}_{N}$ (resp. $\bm{1}_{N}$)
denotes the null (resp. ones) vector of length $N$;
$\mathrm{diag}(\bm{a})$
denotes the diagonal matrix with $\bm{a}$ on the main diagonal; $\lambda_{max}(\bm{A})$ evaluates the highest eigenvalue from the positive-definite matrix $\bm{A}$; $\underline{\bm{a}}$
(resp. $\underline{\bm{A}}$) denotes the augmented vector (resp.
matrix) of $\bm{a}$ (resp. $\bm{A}$), that is $\underline{\bm{a}}\triangleq\left[\begin{array}{cc}
\bm{a}^{T} & \bm{a}^{\dagger}\end{array}\right]^{T}$ (resp. $\underline{\bm{A}}\triangleq\left[\begin{array}{cc}
\bm{A}^{T} & \bm{A}^{\dagger}\end{array}\right]^{T}$); $\Pr(\cdot)$ and $p(\cdot)$ denote probability mass functions (pmfs) and probability density functions (pdfs), while $\Pr(\cdot|\cdot)$ and $p(\cdot|\cdot)$ their corresponding conditional counterparts; $\mathcal{N}_{\mathbb{C}}(\bm{\mu},\bm{\Sigma})$ denotes a proper complex normal distribution with mean vector $\bm{\mu}$ and covariance matrix $\bm{\Sigma}$;
$\mathcal{U}(a,b)$ denotes a uniform distribution with support $(a,b)$;
the symbol $\sim$ means \say{distributed as}.

\section{System Model}\label{Model}
This section describes the DD task for a WSN setup operating with a holographic DF approach.
First, the sensing, local decision and modulation assumptions are stated (Sec.~\ref{subsec:sensing_model}).
Then, the channel model between the sensors and RHS (far-field) and the RHS and the FC (near-field) is described (Sec.~\ref{subsec:channel_model}).
The section ends with the (conditional) statistical characterization of the received signal vector (Sec.~\ref{subsec:stat_char}).

\subsection{Sensing Model \& Modulation Format}\label{subsec:sensing_model}
We consider a distributed binary test of hypotheses, where $K$ sensors are used to discern between the hypotheses in the
set $\mathcal{H}\triangleq\{\mathcal{H}_{0},\mathcal{H}_{1}\}$ (e.g. $\mathcal{H}_{0}/\mathcal{H}_{1}$ may represent the absence/presence of a phenomenon of interest). 
The $k$th sensor, $k\in\mathcal{K}\triangleq\{1,2,\ldots,K\}$, takes a binary local decision $\xi_{k}\in\mathcal{H}$ about the observed phenomenon on the basis of its own measurements. Here we do not make any conditional (given $\mathcal{H}_{i}\in\mathcal{H}$) mutual independence assumption on $\xi_{k}$. 
Each decision $\xi_{k}$
is mapped into $x_{k}\in{\cal X}=\{-1,+1\}$, namely a Binary Phase-Shift Keying (BPSK) modulation: without loss of generality we assume that $b_{k}=\mathcal{H}_{i}$
maps into $x_{k}=(2i-1)$, $i\in\{0,1\}$. 
The quality of the WSN is characterized
by the conditional joint pmfs $P(\bm{x}|\mathcal{H}_{i})$. Also,
we denote $P_{D,k}\triangleq P\left(x_{k}=1|\mathcal{H}_{1}\right)$
and $P_{F,k}\triangleq P\left(x_{k}=1|\mathcal{H}_{0}\right)$ the
probability of detection and false alarm of the $k$th sensor, respectively
(we reasonably assume $P_{D,k}\geq P_{F,k}$). 

\subsection{Channel Model}\label{subsec:channel_model}

\noindent
\textbf{Sensors-RHS links:} sensors communicate their decisions over a wireless flat-fading MAC~\cite{Ciuonzo2012} to an FC equipped with an RHS made of $M$ (reconfigurable) reflecting elements and $N$ external (receive) feeds, each connected to an RF chain.
Let $\bm{H}\in\mathbb{C}^{M\times K}$ ($\bm{h}_{k}\in\mathbb{C}^{M}$ being the $k$th sensor contribution) and $\bm{G}\in\mathbb{C}^{N\times M}$ be the equivalent channels from the WSN to the RHS, and from the RHS to the $N$ feeds, respectively.

We consider a Rician fading model for the link between the $k$th sensor and the RHS, namely
\begin{align}
\bm{h}_{k} & =\sqrt{P(d_{k}^{\mathrm{sen\rightarrow RHS}},\nu)}\,\left(b_{k}\,\bm{h}_{k}^{\mathrm{LoS}}+\sqrt{1-b_{k}^{2}}\,\hat{\bm{h}}_{k}\right)\label{eq: sensor-RIS channel}
\end{align}
where the path loss model considered is
\begin{equation}
P(d,\nu)=\mu\,(d/d_{0})^{-\nu}\label{eq: path_loss_expr_WSN}
\end{equation}
where $\mu$ is the path loss attenuation at the reference distance $d_0$, while $\nu$ denotes the path loss exponent.
In Eq.~\eqref{eq: sensor-RIS channel}, the term $d_{k}^{\mathrm{sen\rightarrow RHS}}$ denotes the $k$th sensor-RHS distance.
\footnote{Since the RHS is assumed in the far-field, the distance is calculated w.r.t. the center of the RHS.}

In Eq.~\eqref{eq: sensor-RIS channel}, the line-of-sight contribution of the aforementioned channel ($\bm{h}_{k}^{\mathrm{LoS}}$) is obtained as:
\begin{equation}
\bm{h}_{k}^{\mathrm{LoS}}=\bm{a}^{\mathrm{upa}}(\theta_{k}^{\mathrm{AoA}},\phi_{k}^{\mathrm{AoA}})\,e^{j\tau_{k}}
\end{equation}
where $\bm{a}^{\mathrm{upa}}(\cdot,\cdot)$ is the steering vector of a uniform planar array, namely:
\begin{equation}
\left[\bm{a}^{\mathrm{upa}}(\theta,\phi)\right]_{m_{x},m_{y}}=\begin{bmatrix}e^{j\frac{2\pi}{\lambda}(m_{x}\,\Delta_{\mathrm{h}}^{\mathrm{rhs}}\sin\theta\cos\phi+m_{y}\,\Delta_{\mathrm{v}}^{\mathrm{rhs}}\sin\theta\sin\phi)}\end{bmatrix}
\end{equation}
where $\lambda$ denotes the wavelength, while $\Delta_{\mathrm{h}}^{\mathrm{rhs}}$ (resp. $\Delta_{\mathrm{v}}^{\mathrm{rhs}}$)
denotes the RHS inter-element spacing along the horizontal (resp. vertical) axis.
The latter term depends on the azimuth-elevation angle-of-arrival pair at the RHS, denoted here with $(\theta_{k}^{\mathrm{AoA}},\phi_{k}^{\mathrm{AoA}})$, since the sensors are assumed to be in the far-field of the RHS.
Differently, the term $\tau_{k}\sim\mathcal{U}(0,2\pi)$ models the different initial delay from each sensor toward the RHS.

Conversely, $\ensuremath{\hat{\bm{h}}_{k}}\sim\ensuremath{\mathcal{N}_{\mathbb{C}}(\bm{0}_{M},\bm{I}_{M})}$ corresponds to the normalized NLOS (scattered) component of the same link.
Last, the scalar term $b_k$ is the Rician factor associated to the link.%
\footnote{We underline that the transmit gain of each sensor is implicitly included in the path loss expression of Eq.~\eqref{eq: path_loss_expr_WSN}, while the different receive aperture/gain experienced by each sensor w.r.t. the RHS is here absorbed in the Rician factor (i.e. via $b_k$).}

\noindent
\textbf{RHS-feeds links:} the channel matrix between the RHS and the $N$ external feeds follows a \emph{near-field} model, based on the well-known (deterministic) \emph{spherical wave} equation.
Specifically, the $(n,m)$th element of $\bm{G}$ equals~\cite{jamali2020intelligent}
\begin{equation}\label{eq:channel_model}
g_{n,m}=\left(\frac{\lambda}{4\pi}\right)\sqrt{\eta\,G_{n,m}^{\mathrm{rhs}}G_{n,m}^{\mathrm{fc}}}\,\frac{\exp(-j\,(2\pi/\lambda)\left\Vert \bm{p}_{n}^{\mathrm{fc}}-\bm{p}_{m}^{\mathrm{rhs}}\right\Vert )}{\left\Vert \bm{p}_{n}^{\mathrm{fc}}-\bm{p}_{m}^{\mathrm{rhs}}\right\Vert }
\end{equation}
where:
\begin{itemize}
\item $\bm{p}_{n}^{\mathrm{fc}}$ and $\bm{p}_{m}^{\mathrm{rhs}}$ denote the 3D position of the $n$th feed and of $m$th RHS element, respectively;
\item $G_{n,m}^{\mathrm{rhs}}$ denotes the transmit (re-radiation) gain of the $m$th RHS element toward the $n$th feed;
\item $G_{n,m}^{\mathrm{fc}}$ characterizes the receive gain at the $n$th antenna for power re-radiated from the $m$th RHS element;
\item $\eta$ underlines the RHS power efficiency in reflecting the impinging waves.
\end{itemize}
The (RHS) re-rediation gain $G_{n,m}^{\mathrm{rhs}}$ has the following form:
\begin{equation}
\ensuremath{G_{n,m}^{\mathrm{rhs}}}=\frac{4\pi}{\lambda^{2}}\,A_{n,m}^{\mathrm{rhs}}\label{eq: radiation_gain_rhs}
\end{equation}
where $A_{n,m}^{\mathrm{rhs}}$ denotes the effective aperture of $m$th RHS element as seen by the $n$th feed.  
Similarly, the (feed) receive gain is:
\begin{equation}
\ensuremath{G_{n,m}^{\mathrm{fc}}}=\frac{4\pi}{\lambda^{2}}\,A_{n,m}^{\mathrm{fc}}\label{eq: FC receive gain}
\end{equation}
where $A_{n,m}^{\mathrm{fc}}$ denotes the effective aperture of the $n$th feed as seen by the $m$th RHS element.

The effective aperture \( A_{n,m}^{\mathrm{rhs}} \) is determined by scaling the maximum aperture—achieved when the element is perfectly aligned in the boresight direction—by a directivity factor \( \rho_{n,m}^{\mathrm{rhs}} \) that accounts for angular misalignment. This gives:  
\begin{equation}
A_{n,m}^{\mathrm{rhs}}=\left(\Delta_{\mathrm{h}}^{\mathrm{rhs}}\Delta_{\mathrm{v}}^{\mathrm{rhs}}\right)\cdot\rho_{n,m}^{\mathrm{rhs}}
\end{equation} 
An analogous relationship holds for the feed element:  
\begin{equation}
A_{n,m}^{\mathrm{fc}}=\left(\Delta_{\mathrm{h}}^{\mathrm{fc}}\Delta_{\mathrm{v}}^{\mathrm{fc}}\right)\cdot\rho_{n,m}^{\mathrm{fc}}
\end{equation}
Notably, when the RHS element is perfectly aligned, its maximum aperture is simply given by its physical area, which is the product of its horizontal and vertical dimensions, \( \Delta_{\mathrm{h}}^{\mathrm{rhs}} \) and \( \Delta_{\mathrm{v}}^{\mathrm{rhs}} \), respectively. A similar consideration applies to the feed element, where the maximum aperture equals the feed element area, computed as \( \Delta_{\mathrm{h}}^{\mathrm{fc}} \Delta_{\mathrm{v}}^{\mathrm{fc}} \).  

In both the gain terms of Eqs.~\eqref{eq: radiation_gain_rhs}~and~\eqref{eq: FC receive gain}, the directivity factors $\rho_{n,m}^{\mathrm{rhs}}$ and $\rho_{n,m}^{\mathrm{fc}}$ are obtained via the functional form $\rho(\theta,\phi)$, which has the following expression~\cite{jamali2020intelligent,ellingson2021path}:
\begin{align}
\rho(\theta,\phi) & \triangleq\begin{cases}
2\,(2q+1)\cos^{2q}(\theta) & \begin{array}{c}
0\leq\theta\leq\pi/2\,\\
0\leq\phi\leq2\pi
\end{array},\\
0 & \mathrm{otherwise}
\end{cases}\label{eq: general directivity pattern}
\end{align}
where $q\geq 0 $ is a real number that determines the directivity, while the multiplication factor $2\,(2q+1)$ ensures energy conservation ($\frac{1}{4\pi}\int_{\Omega}\rho(\theta,\phi)\,d\Omega=1$).
The same term also represents the value achieved in the direction of maximum gain~\cite{tang2020wireless}.
A few relevant pattern gain examples are the \emph{isotropic (half-space)} pattern
\begin{align}
\rho(\theta,\phi) & \triangleq\begin{cases}
2 & \begin{array}{c}
0\leq\theta\leq\pi/2\,\\
0\leq\phi\leq2\pi
\end{array},\\
0 & \mathrm{otherwise}
\end{cases}
\end{align}
and the \emph{cosine pattern}~\cite{tang2022path}:
\begin{align}
\rho(\theta,\phi) & \triangleq\begin{cases}
4\,\cos(\theta) & \begin{array}{c}
0\leq\theta\leq\pi/2\,\\
0\leq\phi\leq2\pi
\end{array},\\
0 & \mathrm{otherwise}
\end{cases}
\end{align}
The aforementioned pattern expressions can be calculated as~\cite{abrardo2021intelligent,DegliEsposti2022,feng2023near}:
\begin{align}
\rho_{n,m}^{\mathrm{rhs}} & =2\,(2q+1)\,\cos^{2q}(\theta_{n,m}^{\mathrm{rhs}})\\
\rho_{n,m}^{\mathrm{fc}} & =2\,(2q+1)\,\cos^{2q}(\theta_{n,m}^{\mathrm{fc}})
\end{align}
where 
\begin{align}
\cos(\theta_{n,m}^{\mathrm{rhs}})= & \frac{(\bm{p}_{n}^{\mathrm{fc}}-\bm{p}_{m}^{\mathrm{rhs}})^{T}\,\bm{u}^{\mathrm{rhs}}}{\left\Vert \bm{p}_{n}^{\mathrm{fc}}-\bm{p}_{m}^{\mathrm{rhs}}\right\Vert }\,\\
\cos(\theta_{n,m}^{\mathrm{fc}})= & \frac{(\bar{\bm{p}}^{\mathrm{rhs}}-\bm{p}_{n}^{\mathrm{fc}})^{T}\,\left(\bm{p}_{m}^{\mathrm{rhs}}-\bm{p}_{n}^{\mathrm{fc}}\right)}{\left\Vert \bar{\bm{p}}^{\mathrm{rhs}}-\bm{p}_{n}^{\mathrm{fc}}\right\Vert \left\Vert \bar{\bm{p}}^{\mathrm{rhs}}-\bm{p}_{n}^{\mathrm{fc}}\right\Vert }
\end{align}
In other terms, $\cos(\theta_{n,m}^{\mathrm{rhs}})$ represents the cosine angle between the RHS element boresight direction and the vector denoting the reflection direction of the $m$th RHS element toward $n$th receive feed.  
Conversely, $\cos(\theta_{n,m}^{\mathrm{fc}})$ represents the cosine angle between the same reflection direction (but in the coordinate system of $n$th feed) and the vector of maximum receive gain. 
In this work, we assume that $N$ receive antennas are aligned with their maximum gain directions towards the geometrical center of the RHS ($\bar{\bm{p}}^{\mathrm{rhs}}$), namely a full-illumination configuration~\cite{jamali2020intelligent}.

In compact form, the received signal vector $\bm{y}\in\mathbb{C}^{N}$ at the FC can be written as:
\begin{align}
\bm{y}= & \left(\bm{G}\bm{\Theta}\bm{H}\right)\bm{D}_{\alpha}\,\bm{x}+\bm{w}=\bm{H}^{e}(\bm{\Theta})\,\bm{D}_{\alpha}\,\bm{x}+\bm{w}\label{eq: signal_model}
\end{align}
where $\bm{x}\in\mathcal{X}^{K}$ and $\bm{w}\sim\mathcal{N}_{\mathbb{C}}(\bm{0}_{N},\ensuremath{\sigma_{w}^{2}\bm{I}_{N})}$ are the transmitted signal and noise vectors, respectively. 
In Eq.~\eqref{eq: signal_model}, the diagonal matrix $\bm{\Theta}=\mathrm{diag}(e^{j\varphi_{1}},\ldots,e^{j\varphi_{M}})$, $0\leq\varphi_{m}<2\pi$, $\forall m=1,\ldots M$, collects the phase-shifts of the RHS.
Conversely, the matrix $\bm{D}_{\alpha}=\mathrm{diag}(\alpha_{1},\ldots,\alpha_{K})$,
with $\alpha_{k}\in\mathbb{R}^{+},\,\forall k \in \mathcal{K}$, accounts for unequal transmit energy from the sensors.
Last, we have defined $\bm{H}^{e}(\bm{\Theta})\,\triangleq\left(\bm{G}\bm{\Theta}\bm{H}\right)$ for compactness.

\noindent
\textbf{Remarks on channel knowledge:} in this work, we assume that \emph{both} channel matrices $\bm{G}$ and $\bm{H}$ are available for the proposed fusion rule \& RHS design.
While this is a common assumption for the former channel because a fixed and precise RHS-feeds configuration can be ensured during manufacturing, the channel $\bm{H}$ can be obtained by existing channel estimation techniques (by exploiting the knowledge of $\bm{G}$).

\noindent
\textbf{Remarks on conventional DF:} in this work, the proposed holographic DF system is compared against a (conventional) \emph{fully-digital} MIMO architecture at the FC~\cite{Ciuonzo2012,ciuonzo2015} equipped with $N_{dig}$ antennas and RF chains, arranged in a planar array whose geometric center is the same as $\bar{\bm{p}}^{\mathrm{rhs}}$ and same planar alignment as the RHS.
In the latter case, the channel model $\ensuremath{\bm{H}\in\mathbb{C}^{N_{dig}\times K}}$ between the sensors and the fully-digital array is a Rician channel, consistent with the holographic scenario, using \emph{identical Rician factors} $b_k$'s.
As a result, the system model in such a case is obtained by substituting $\bm{H}^{e}(\bm{\Theta})$ with $\bm{H}$ in Eq.~\eqref{eq: signal_model}. 

\subsection{Statistical Characterization}\label{subsec:stat_char}
Based on the aforementioned assumptions, the vector $\bm{y}|\mathcal{H}_{i}$ has the following (hypothesis-conditional) statistical characterization at the second order:
\begin{align}
\ensuremath{\mathbb{E}\{\bm{y}|\mathcal{H}_{i}\}} & =\bm{H}^{e}(\bm{\Theta})\,\bm{D}_{\alpha}\,(2\,\bm{\rho}_{i}-\bm{1}_{K})\label{eq:2nd_order_char}\\
\ensuremath{\mathrm{Cov}(\bm{y}|\mathcal{H}_{i})} & \ensuremath{=}\bm{H}^{e}(\bm{\Theta})\,\bm{D}_{\alpha}\,\mathrm{Cov}\left(\bm{x}|\mathcal{H}_{i}\right)\,\bm{D}_{\alpha}\,\bm{H}^{e}(\bm{\Theta})^{\dagger}+\sigma_{w}^{2}\,\bm{I}_{N}\nonumber \\
\mathrm{PCov}(\bm{y}|\mathcal{H}_{i}) & =\bm{H}^{e}(\bm{\Theta})\,\bm{D}_{\alpha}\,\mathrm{Cov}(\bm{x}|\mathcal{H}_{i})\,\bm{D}_{\alpha}\,\bm{H}^{e}(\bm{\Theta})^{T}\nonumber 
\end{align}
where $\mathrm{Cov}\left(\bm{x}|\mathcal{H}_{i}\right)$ denotes the $\mathcal{H}_{i}$-conditional covariance matrix of the decision vector $\bm{x}$. Additionally, the column vectors $\ensuremath{\bm{\rho}_{1}\triangleq[P_{D,1}\cdots P_{D,K}]^{T}}$
and $\bm{\rho}_{0}\triangleq[P_{F,1}\cdots P_{F,K}]^{T}$ collect the detection and false-alarm probabilities of all the sensors,
respectively.
It is worth underlining that, given the improper nature of the vector $\bm{y}|\mathcal{H}_{i}$, a complete second-order characterization must include the pseudo-covariance $\mathrm{PCov}(\bm{y}|\mathcal{H}_{i})$.

\section{Joint Fusion Rule and RHS design Problem Formulation}\label{sec_fusion_rule_RIS}
The objective of this paper is to formulate a feasible design strategy to achieve high (goal-oriented) performance, consisting in the joint design of a fusion rule statistic $\Lambda = \mathcal{F}(\bm{y})$ and of the RHS matrix $\bm{\Theta}$. 
Accordingly, we first recall the optimal fusion rule (when the RHS configuration is \emph{fixed}) and underline the impossibility of pursuing this strategy (Sec.~\ref{subsec: optimal_LLR}).
Then, we introduce two considered design strategies (FuC and IS) when the fusion rule is constrained to be a widely-linear processing of the received signal vector (Sec.~\ref{subsec: WLlinear_design}).

\subsection{Optimal Fusion Rule}\label{subsec: optimal_LLR}
 
The optimal test~\cite{Kay1998} for this problem is
\begin{equation}
\left\{ \Lambda_{\mathrm{opt}}\triangleq\ln\left[\frac{p(\bm{y}|\mathcal{H}_{1})}{p(\bm{y}|\mathcal{H}_{0})}\right]\right\} \begin{array}{c}
{\scriptstyle \hat{\mathcal{H}}=\mathcal{H}_{1}}\\
\gtrless\\
{\scriptstyle \hat{\mathcal{H}}=\mathcal{H}_{0}}
\end{array}\gamma\label{eq:neyman_pearson_test}
\end{equation}
where $\hat{\mathcal{H}}$, $\Lambda_{\mathrm{opt}}$ and $\gamma$
denote the estimated hypothesis, the LLR and the threshold which the
LLR is compared to.
The threshold $\gamma$ is usually determined to assure a fixed system false-alarm rate or to minimize the probability of error~\cite{Kay1998}.
Exploiting the independence\footnote{Indeed the directed triple formed by hypothesis, the transmitted-signal
vector and the received-signal vector satisfies the Markov property.} of $\bm{y}$ from $\mathcal{H}_{i}$, given $\bm{x}$, an explicit
expression of the LLR in Eq.~(\ref{eq:neyman_pearson_test}) is obtained as 
\begin{gather}
\Lambda_{\mathrm{opt}}=\ln\left[\frac{\sum_{\bm{x}\in{\cal X}^{K}}p(\bm{y}|\bm{x})P(\bm{x}|\mathcal{H}_{1})}{\sum_{\bm{x}\in{\cal X}^{K}}p(\bm{y}|\bm{x})P(\bm{x}|\mathcal{H}_{0})}\right]\label{eq:LLR_RIS}\\
=\ln\left[\frac{\sum_{\bm{x}\in{\cal X}^{K}}\exp\left(-\frac{\bm{\|y}-\bm{H}^{e}(\bm{\Theta})\bm{D}_{\alpha}\bm{x}\|^{2}}{\sigma_{w}^{2}}\right)\Pr(\bm{x}|\mathcal{H}_{1})}{\sum_{\bm{x}\in{\cal X}^{K}}\exp\left(-\frac{\bm{\|y}-\bm{H}^{e}(\bm{\Theta})\bm{D}_{\alpha}\bm{x}\|^{2}}{\sigma_{w}^{2}}\right)\Pr(\bm{x}|\mathcal{H}_{0})}\right]\nonumber 
\end{gather}
Unfortunately,  the LLR requires a computational complexity which scales with $2^{K}$ and also does not lend itself to a tractable analysis of its performance.

\subsection{Deflection-based Widely-Linear \\ RHS-based Fusion Rule Design}\label{subsec: WLlinear_design}

The latter observation still applies even if a deflection measure is taken as the relevant evaluation metric:
\begin{gather}
D_{i}\left(\Lambda\right)\triangleq\left(\mathbb{E}\{\Lambda|\mathcal{H}_{1}\}-\mathbb{E}\{\Lambda|\mathcal{H}_{0}\}\right)^{2}\,/\,\mathrm{var}\{\Lambda|\mathcal{H}_{i}\}
\end{gather}
where $D_{0}(\cdot)$ and $D_{1}(\cdot)$
correspond to the normal~\cite{Picinbono1995} and modified~\cite{Quan2008}
deflections, respectively.
Accordingly, we pursue a \emph{simplified approach} described in what follows.

First of all, deflection metrics represent a flexible design tool when the processing at the FC is constrained to be a Widely-Linear (WL) fusion statistic
\begin{equation}
\Lambda_{\mathrm{wl}}=\underline{\bm{a}}^{\dagger}\bm{\underline{y}}\label{eq: WL fusion statistic}
\end{equation}
The statistic $\Lambda_{\mathrm{wl}}$ is then used to implement the hypothesis test similarly to Eq.~\eqref{eq:neyman_pearson_test}.
The WL approach is motivated by reduced
complexity and $\bm{y}|\mathcal{H}_{i}$ being an \emph{improper} complex-valued random vector, that is $\mathrm{PCov}(\bm{y}|\mathcal{H}_{i})\neq\bm{O}_{N\times N}$ (see third line of Eq.~\eqref{eq:2nd_order_char}).
In that case, the expression of the generic deflection measure simplifies as:
\begin{gather}
D_{i}(\Lambda_{\mathrm{wl}})\triangleq\frac{\left(\underbar{\ensuremath{\bm{a}}}^{\dagger}\left(\mathbb{E}\{\text{\ensuremath{\underbar{\ensuremath{\bm{y}}}}}|\mathcal{H}_{1}\}-\mathbb{E}\{\text{\ensuremath{\underbar{\ensuremath{\bm{y}}}}}|\mathcal{H}_{0}\}\right)\right)^{2}}{\underbar{\ensuremath{\bm{a}}}^{\dagger}\mathrm{Cov}(\text{\ensuremath{\underbar{\ensuremath{\bm{y}}}}}|\mathcal{H}_{i})\,\underbar{\ensuremath{\bm{a}}}}
\end{gather}
Indeed, WL fusion rules have been reported to provide appealing performance in similar WSN contexts (e.g.~\cite{ciuonzo2015}).
 
In what follows, we consider \emph{two design strategies}: ($i$) one based on the fully-complete second order characterization reported in Eq.~\eqref{eq:2nd_order_char}; ($ii$) a simplified design based on the well-known IS assumption~\cite{lei2010coherent,ciuonzo2015}.
The latter hypothesizes perfect sensing at the design stage, namely $\Pr(\bm{x}=\bm{1}_{K}|\mathcal{H}_{1})=\Pr(\bm{x}=-\bm{1}_{K}|\mathcal{H}_{0})=1$.
As a result, an IS-based design strategy requires lower system knowledge, since \emph{WSN (local) decision performance is not needed}.

First, in the case of full characterization (FuC), the deflection measures can be rewritten as:
\begin{equation}
D_{\mathrm{\mathrm{FuC},}i}\left(\text{\ensuremath{\underbar{\ensuremath{\bm{a}}}}},\bm{\Theta}\right)=\frac{4\cdot\left(\underbar{\ensuremath{\bm{a}}}^{\dagger}\left[\underline{\bm{H}}^{e}(\bm{\Theta})\,\bm{D}_{\alpha}\,\ensuremath{\bm{\rho}_{10}}\right]\right)^{2}}{\underbar{\ensuremath{\bm{a}}}^{\dagger}\mathrm{Cov}(\text{\ensuremath{\underbar{\ensuremath{\bm{y}}}}}|\mathcal{H}_{i})\,\underbar{\ensuremath{\bm{a}}}}\label{eq: deflection_WL}
\end{equation}
where $\bm{\rho}_{10}\triangleq\,(\bm{\rho}_{1}-\bm{\rho}_{0})$, while the \emph{augmented covariance} has the following expression:
\begin{equation}
\mathrm{Cov}(\text{\ensuremath{\underbar{\ensuremath{\bm{y}}}}}|\mathcal{H}_{i})=\underline{\bm{H}}^{e}(\bm{\Theta})\,\bm{D}_{\alpha}\,\mathrm{Cov}(\bm{x}|\mathcal{H}_{i})\,\bm{D}_{\alpha}\,\underline{\bm{H}}^{e}(\bm{\Theta})^{\dagger}+\sigma_{w}^{2}\,\bm{I}_{2N}\label{eq: augmented covariance rx signal vector}
\end{equation}
Differently, based on the IS assumption, the second-order statistical characterization of $\bm{y}|\mathcal{H}_{i}$ simplifies (compare it with Eq.~\eqref{eq:2nd_order_char}) as:
\begin{eqnarray}
\mathbb{E}\{\bm{y}|\mathcal{H}_{i}\} & = & \bm{H}^{e}(\bm{\Theta})\,\bm{D}_{\alpha}\,(2i-1)\bm{1}_{K}\label{eq: 2nd order char IS}\\
\mathrm{Cov}(\bm{y}|\mathcal{H}_{i}) & = & \sigma_{w}^{2}\,\bm{I}_{N}\nonumber \\
\mathrm{PCov}(\bm{y}|\mathcal{H}_{i}) & = & \bm{O}_{N\times N}\nonumber 
\end{eqnarray}
Accordingly, in such a case both the deflection measures simplify into the following \emph{unique} metric~\cite{mudkey2022wireless}:
\begin{equation}
D_{\mathrm{IS}}\left(\text{\ensuremath{\underbar{\ensuremath{\bm{a}}}}},\bm{\Theta}\right)=\frac{4}{\sigma_{w}^{2}}\,\frac{\left(\underbar{\ensuremath{\bm{a}}}^{\dagger}\left[\underline{\bm{H}}^{e}(\bm{\Theta})\,\bm{D}_{\alpha}\,\bm{1}_{k}\right]\right)^{2}}{\underbar{\ensuremath{\bm{a}}}^{\dagger}\underbar{\ensuremath{\bm{a}}}}\label{eq: deflection_IS}
\end{equation}
The goal of this work is to maximize the deflection measures in Eqs.~\eqref{eq: deflection_WL}~and~\eqref{eq: deflection_IS} (i.e. under FuC and IS assumptions, respectively) by \emph{jointly} optimizing the WL vector $\underbar{\ensuremath{\bm{a}}}$ and the matrix $\bm{\Theta}$ determining the fusion rule.

Hence, the resulting optimization problems are formulated as:
\begin{equation}
\mathcal{P}_{\mathrm{\mathrm{FuC}},i}\,:\begin{array}{c}
\underset{\underbar{\ensuremath{\bm{a}}}\,,\bm{\Theta}}{\mathrm{maximize}}\frac{\left(\underbar{\ensuremath{\bm{a}}}^{\dagger}\left[\underline{\bm{H}}^{e}(\bm{\Theta})\,\bm{D}_{\alpha}\,\ensuremath{\bm{\rho}_{10}}\right]\right)^{2}}{\underbar{\ensuremath{\bm{a}}}^{\dagger}\mathrm{Cov}(\text{\ensuremath{\underbar{\ensuremath{\bm{y}}}}}|\mathcal{H}_{1})\,\underbar{\ensuremath{\bm{a}}}}\qquad\\
\mathrm{subject\,to}\begin{array}{c}
\left\Vert \underbar{\ensuremath{\bm{a}}}\right\Vert =1\\
\bm{\Theta}=\mathrm{diag}(e^{j\varphi_{1}},\ldots,e^{j\varphi_{M}})
\end{array}
\end{array}
\end{equation}
\begin{equation}
\mathcal{P}_{\mathrm{IS}}\,:\begin{array}{c}
\underset{\underbar{\ensuremath{\bm{a}}}\,,\bm{\Theta}}{\mathrm{maximize}}\frac{\left(\underbar{\ensuremath{\bm{a}}}^{\dagger}\left[\underline{\bm{H}}^{e}(\bm{\Theta})\,\bm{D}_{\alpha}\,\bm{1}_{k}\right]\right)^{2}}{\underbar{\ensuremath{\bm{a}}}^{\dagger}\underbar{\ensuremath{\bm{a}}}}\qquad\\
\mathrm{subject\,to}\begin{array}{c}
\left\Vert \underbar{\ensuremath{\bm{a}}}\right\Vert =1\\
\bm{\Theta}=\mathrm{diag}(e^{j\varphi_{1}},\ldots,e^{j\varphi_{M}})
\end{array}
\end{array}
\end{equation}
Unlike conventional channel-aware DD schemes that employ fully-digital arrays at the FC, the introduction of the RHS imposes an additional structural constraint: each diagonal entry in the matrix $\bm{\Theta}$ must have a unit modulus.
This constraint, inherently non-convex, further exacerbates the challenge posed by the already non-convex objective function. As a result, both optimization problems $\mathcal{P}_{\mathrm{\mathrm{FuC}},i}$ and $\mathcal{P}_{\mathrm{IS}}$ remain non-convex, making their solution more intricate.

\section{Proposed Solution: AO-based approach}\label{sec:design_solution}
In this paper, we resort to an AO approach for solving $\mathcal{P}_{\mathrm{FuC},i}$ and $\mathcal{P}_{\mathrm{IS}}$ efficiently.
Specifically, we describe an optimization method based on AO which alternates between the maximization of $\underbar{\ensuremath{\bm{a}}}$ and $\bm{\Theta}$ (Secs.~\ref{subsec: fusion_rule_design}~and~\ref{subsec:RIS_design}, respectively). 
AO (being a special case of block-coordinate descent) has been shown to be a widely applicable and empirically successful approach in many applications, and typically leads to a sub-optimal solution for nonconvex problems.
In what follows, we describe the AO-based approach for both FuC and IS cases.
The section ends with the complexity analysis of the design and knowledge requirements (Sec.~\ref{subsec:complexity}).

\subsection{Step (A): Fusion Rule Design} \label{subsec: fusion_rule_design}
We first focus on the  optimization of the WL vector $\underbar{\ensuremath{\bm{a}}}$ for a \emph{fixed} matrix $\bm{\Theta}_{\mathrm{fix}}$.
Accordingly, the WL vector design problem in the FuC case is given by:
\begin{equation}
\mathcal{P}_{\mathrm{FuC},i}^{(\mathrm{A)}}\,:\underset{\left\Vert \underbar{\ensuremath{\bm{a}}}\right\Vert =1}{\mathrm{maximize}}\;\frac{\left(\underbar{\ensuremath{\bm{a}}}^{\dagger}\left[\underline{\bm{H}}^{e}(\bm{\Theta}_{\mathrm{fix}})\,\bm{D}_{\alpha}\,\ensuremath{\bm{\rho}_{10}}\right]\right)^{2}}{\underbar{\ensuremath{\bm{a}}}^{\dagger}\mathrm{Cov}_{\bm{\Theta}_{\mathrm{fix}}}(\text{\ensuremath{\underbar{\ensuremath{\bm{y}}}}}|\mathcal{H}_{i})\,\underbar{\ensuremath{\bm{a}}}}
\end{equation}
where $\mathrm{Cov}_{\bm{\Theta}_{\mathrm{fix}}}(\text{\ensuremath{\underbar{\ensuremath{\bm{y}}}}}|\mathcal{H}_{i})$
means that $\bm{\Theta}\rightarrow\bm{\Theta}_{\mathrm{fix}}$ in Eq.~\eqref{eq: augmented covariance rx signal vector}.
The optimal value of $\underbar{\ensuremath{\bm{a}}}$ in $\mathcal{P}_{\mathrm{FuC},i}^{(\mathrm{A)}}$ is the vector attaining the equality in the \emph{Cauchy-Schwarz inequality}~\cite{ciuonzo2015}:
\begin{equation}
\text{\ensuremath{\underbar{\ensuremath{\bm{a}}}}}_{\mathrm{FuC,}i}^{\star}(\bm{\Theta}_{\mathrm{fix}})=\frac{\mathrm{Cov}_{\bm{\Theta}_{\mathrm{fix}}}(\text{\ensuremath{\underbar{\ensuremath{\bm{y}}}}}|\mathcal{H}_{i})^{-1}\,\underline{\bm{H}}^{\mathrm{e}}(\bm{\Theta}_{\mathrm{fix}})\,\bm{D}_{\alpha}\,\ensuremath{\bm{\rho}_{10}}}{\left\Vert \mathrm{Cov}_{\bm{\Theta}_{\mathrm{fix}}}(\text{\ensuremath{\underbar{\ensuremath{\bm{y}}}}}|\mathcal{H}_{i})^{-1}\,\underline{\bm{H}}^{\mathrm{e}}(\bm{\Theta}_{\mathrm{fix}})\,\bm{D}_{\alpha}\,\ensuremath{\bm{\rho}_{10}}\right\Vert }\label{eq: WL fusion beamformer Step A - FC}
\end{equation}
It is worth noticing that under the IS assumption, the computation of the optimal value does not require whitening by the (augmented) received data covariance, i.e. it simplifies into
\begin{equation}
\text{\ensuremath{\underbar{\ensuremath{\bm{a}}}}}_{\mathrm{IS}}^{\star}(\bm{\Theta}_{\mathrm{fix}})=\frac{\underline{\bm{H}}^{\mathrm{e}}(\bm{\Theta}_{\mathrm{fix}})\,\bm{D}_{\alpha}\,\bm{1}_{k}}{\left\Vert \underline{\bm{H}}^{\mathrm{e}}(\bm{\Theta}_{\mathrm{fix}})\,\bm{D}_{\alpha}\,\bm{1}_{k}\right\Vert }\label{eq: WL fusion beamformer Step A - IS}
\end{equation}
\noindent
\textbf{Remarks on conventional DF:} it is worth noticing that in the case of a fully-digital array (i.e. $\bm{H}^{e}(\bm{\Theta}_{\mathrm{fix}})$ replaced by $\bm{H}$), the corresponding FuC and IS fusion rules~\cite{ciuonzo2015} are obtained by leveraging Eqs.~\eqref{eq: WL fusion beamformer Step A - FC}~and~\eqref{eq: WL fusion beamformer Step A - IS}, respectively, in a single-step fashion (as there is no dependence on $\bm{\Theta}$). 

\subsection{Step (B): RHS Design} \label{subsec:RIS_design}
We then focus on the optimization of the matrix $\bm{\Theta}$ for a \emph{fixed} WL vector  $\text{\ensuremath{\underbar{\ensuremath{\bm{a}}}}}_{\mathrm{fix}}$.
The corresponding optimization problem is given by:
\begin{equation}
\mathcal{P}_{\mathrm{FuC},i}^{(\mathrm{B)}}\,:\begin{array}{c}
\underset{\bm{\Theta}}{\mathrm{maximize}}\frac{\left(\underbar{\ensuremath{\bm{a}}}_{\mathrm{fix}}^{\dagger}\left[\underline{\bm{H}}^{e}(\bm{\Theta})\,\bm{D}_{\alpha}\,\ensuremath{\bm{\rho}_{10}}\right]\right)^{2}}{\underbar{\ensuremath{\bm{a}}}_{\mathrm{fix}}^{\dagger}\mathrm{Cov}_{\bm{\Theta}}(\text{\ensuremath{\underbar{\ensuremath{\bm{y}}}}}|\mathcal{H}_{1})\underbar{\ensuremath{\bm{a}}}_{\mathrm{fix}}}\\
\mathrm{subject\,to}\;\bm{\Theta}=\mathrm{diag}(e^{j\varphi_{1}},\ldots,e^{j\varphi_{M}})
\end{array}
\end{equation}
After some manipulations, the deflection in Eq.~\eqref{eq: deflection_WL} can be recast as follows (isolating the dependence on RHS matrix):
\begin{equation}
D_{\mathrm{FuC},i}\left(\text{\ensuremath{\underbar{\ensuremath{\bm{a}}}}}_{\mathrm{fix}},\bm{\Theta}\right)=\,\frac{\underbar{\ensuremath{\bm{\theta}}}^{\dagger}\bm{\Xi}\left(\text{\ensuremath{\underbar{\ensuremath{\bm{a}}}}}_{\mathrm{fix}}\right)\underbar{\ensuremath{\bm{\theta}}}}{\underbar{\ensuremath{\bm{\theta}}}^{\dagger}\bm{\Psi}\left(\text{\ensuremath{\underbar{\ensuremath{\bm{a}}}}}_{\mathrm{fix}}\right)\underbar{\ensuremath{\bm{\theta}}}}\label{eq: D_full_i}
\end{equation}
In the above expression, the matrix $\bm{\Xi}\left(\text{\ensuremath{\underbar{\ensuremath{\bm{a}}}}}_{\mathrm{fix}}\right)$ is defined as
\begin{equation}
\bm{\Xi}\left(\text{\ensuremath{\underbar{\ensuremath{\bm{a}}}}}_{\mathrm{fix}}\right)\triangleq\left(\bm{N}^{\dagger}\text{\ensuremath{\underbar{\ensuremath{\bm{a}}}}}_{\mathrm{fix}}\right)\left(\bm{N}^{\dagger}\text{\ensuremath{\underbar{\ensuremath{\bm{a}}}}}_{\mathrm{fix}}\right)^{\dagger}\label{eq:Chi}
\end{equation}
where the auxiliary matrix term 
$\bm{N}\in\mathbb{C}^{2N\times2M}$ has the following explicit expression:
\begin{equation}
\bm{N}\triangleq\begin{bmatrix}\bm{N}_{r} & \bm{O}_{N\times M}\\
\bm{O}_{N\times M} & \bm{N}_{r}^{*}
\end{bmatrix}\label{eq: ENNE}
\end{equation}
where $\bm{N}_{r}\triangleq\bm{G}\,\mathrm{diag}(\bm{H}\,\bm{D}_{\alpha}\,\bm{\rho}_{10})$, being a $N\times M$ matrix.

Differently, the matrix $\bm{\Psi}\left(\text{\ensuremath{\underbar{\ensuremath{\bm{a}}}}}_{\mathrm{fix}}\right)$ at the denominator of Eq.~\eqref{eq: D_full_i} is defined as follows:
\begin{gather}
\ensuremath{\bm{\Psi}\left(\text{\ensuremath{\underbar{\ensuremath{\bm{a}}}}}_{\mathrm{fix}}\right)}\triangleq\\
\bm{\Delta}_{0}^{\dagger}\,\bm{D}_{\alpha}\,\mathrm{Cov}(\bm{x}|\mathcal{H}_{i})\,\bm{D}_{\alpha}\,\bm{\Delta}_{0}+\frac{\sigma_{w}^{2}}{2M}\,\left\Vert \text{\ensuremath{\underbar{\ensuremath{\bm{a}}}}}_{\mathrm{fix}}\right\Vert ^{2}\bm{I}_{2M}\nonumber 
\end{gather}
where 
\begin{equation}
\bm{\Delta}_{0}\triangleq\begin{bmatrix}\bm{D}_{r}^{*} & \bm{D}_{r}\end{bmatrix}\,,
\end{equation}
and the following auxiliary definition has been used $\bm{D}_{r}\triangleq\bm{H}^{\dagger}\,\mathrm{diag}(\bm{G}^{\dagger}\,\bm{a}_{\mathrm{fix}})$.
Accordingly, $\bm{D}_{r}$ has size $K \times M$.

Therefore, the optimization problem $\mathcal{P}_{\mathrm{FuC},i}^{(\mathrm{B)}}$ can be recast as:
\begin{equation}
\bar{\mathcal{P}}_{\mathrm{FuC},i}^{(\mathrm{B)}}\,:\begin{array}{c}
\underset{\bm{\theta}}{\mathrm{maximize}}\quad g_{\mathrm{FuC},i}(\bm{\theta})=\frac{\underbar{\ensuremath{\bm{\theta}}}^{\dagger}\,\bm{\Xi}\left(\text{\ensuremath{\underbar{\ensuremath{\bm{a}}}}}_{\mathrm{fix}}\right)\,\underbar{\ensuremath{\bm{\theta}}}}{\underbar{\ensuremath{\bm{\theta}}}^{\dagger}\,\bm{\Psi}\left(\text{\ensuremath{\underbar{\ensuremath{\bm{a}}}}}_{\mathrm{fix}}\right)\,\underbar{\ensuremath{\bm{\theta}}}}\\
\mathrm{subject\,to\quad}\left\{ |\theta_{m}|=1\right\} _{m=1}^{M}
\end{array}
\end{equation}
Differently, in the case of the IS-based design the associated objective, here denoted by $g_{\mathrm{IS}}(\bm{\theta})$, specializes (and simplifies) into:
\begin{equation}
\bar{\mathcal{P}}_{\mathrm{IS}}^{(\mathrm{B)}}\,:\begin{array}{c}
\underset{\bm{\theta}}{\mathrm{maximize}}\quad g_{\mathrm{IS}}(\bm{\theta})=\frac{\underbar{\ensuremath{\bm{\theta}}}^{\dagger}\tilde{\bm{\Xi}}\left(\text{\ensuremath{\underbar{\ensuremath{\bm{a}}}}}_{\mathrm{fix}}\right)\underbar{\ensuremath{\bm{\theta}}}}{\left\Vert \text{\ensuremath{\underbar{\ensuremath{\bm{a}}}}}_{\mathrm{fix}}\right\Vert ^{2}}\\
\mathrm{subject\,to\quad}\left\{ |\theta_{m}|=1\right\} _{m=1}^{M},
\end{array}
\end{equation}
where $\tilde{\bm{\Xi}}\left(\text{\ensuremath{\underbar{\ensuremath{\bm{a}}}}}_{\mathrm{fix}}\right)$
is obtained by replacing $\bm{\rho}_{10}\rightarrow\bm{1}_{K}$ into
$\bm{N}_{r}$ before using Eqs.~\eqref{eq: ENNE}~and~\eqref{eq:Chi}.
Due to modulus constraint of RHS elements, the problem is \emph{non-convex} in both cases. 
Hence, to avoid cumbersome optimizations at the FC, we propose to solve $\bar{\mathcal{P}}_{\mathrm{FuC},i}^{(\mathrm{B)}}$ and $\bar{\mathcal{P}}_{\mathrm{IS}}^{(\mathrm{B)}}$ via the MM technique%
\footnote{Actually, we exploit the \say{dual} minorization-maximization formulation.}, which is briefly recalled in what follows~\cite{sun2016majorization}.

\noindent
\textbf{Refresher on MM:} Denoting the value of $\bm{\theta}$ in the $\ell$th iteration of the AO by $\bm{\theta}_{(\ell)}^{\star}$, we construct a lower bound on the objective function $g(\bm{\theta})$ that touches the objective function at point $\bm{\theta}$, denoted as $f(\bm{\theta}|\bm{\theta}_{(\ell)}^{\star})$.
We adopt this lower bound as a surrogate objective function, and the maximizer of this surrogate objective function is then taken as the value of $\bm{\theta}$ in the next iteration of the AO, i.e., $\bm{\theta}_{(\ell+1)}^{\star}$. In this way, the objective value is monotonically increasing from one iteration to the next, i.e., $g(\bm{\theta}_{(\ell+1)}^{\star})\geq g(\bm{\theta}_{(\ell)}^{\star})$ and we have first-order optimality. The key to the success of MM lies in constructing a surrogate objective function $f(\bm{\theta}|\bm{\theta}_{(\ell)}^{\star})$
for which the maximizer $\bm{\theta}_{(\ell+1)}^{\star}$ is easy to find.
For the phase-shift matrix optimization problems $\bar{\mathcal{P}}_{\mathrm{FuC},i}^{(\mathrm{B)}}$ and $\bar{\mathcal{P}}_{\mathrm{IS}}^{(\mathrm{B)}}$, surrogate objective functions are derived in what follows.

\noindent
\textbf{Definition of MM surrogate objective:}
Leveraging the convexity of $\underbar{\ensuremath{\bm{\theta}}}^{\dagger}\,\bm{\Xi}\left(\text{\ensuremath{\underbar{\ensuremath{\bm{a}}}}}_{\mathrm{fix}}\right)\,\underbar{\ensuremath{\bm{\theta}}}\,/\,c$ in $\{\bm{\theta},c\}$, the  objective function $g_{\mathrm{FuC},i}(\bm{\theta})$ can be \emph{minorized} as:
\begin{gather}
g_{\mathrm{FuC},i}(\bm{\theta})=\frac{\underbar{\ensuremath{\bm{\theta}}}^{\dagger}\,\bm{\Xi}\left(\text{\ensuremath{\underbar{\ensuremath{\bm{a}}}}}_{\mathrm{fix}}\right)\,\underbar{\ensuremath{\bm{\theta}}}}{\underbar{\ensuremath{\bm{\theta}}}^{\dagger}\,\bm{\Psi}\left(\text{\ensuremath{\underbar{\ensuremath{\bm{a}}}}}_{\mathrm{fix}}\right)\,\underbar{\ensuremath{\bm{\theta}}}}\geq\nonumber \\
f_{\mathrm{FuC},i}(\bm{\theta}|\bm{\theta}_{(\ell)}^{\star})=2\frac{\Re\left\{ \left(\underbar{\ensuremath{\bm{\theta}}}_{(\ell)}^{\star}\right)^{\dagger}\,\bm{\Xi}\left(\text{\ensuremath{\underbar{\ensuremath{\bm{a}}}}}_{\mathrm{fix}}\right)\,\underbar{\ensuremath{\bm{\theta}}}\right\} }{\left(\underbar{\ensuremath{\bm{\theta}}}_{(\ell)}^{\star}\right)^{\dagger}\bm{\Psi}\left(\text{\ensuremath{\underbar{\ensuremath{\bm{a}}}}}_{\mathrm{fix}}\right)\,\underbar{\ensuremath{\bm{\theta}}}}\nonumber \\
-\frac{\left(\underbar{\ensuremath{\bm{\theta}}}_{(\ell)}^{\star}\right)^{\dagger}\,\bm{\Xi}\left(\text{\ensuremath{\underbar{\ensuremath{\bm{a}}}}}_{\mathrm{fix}}\right)\,\underbar{\ensuremath{\bm{\theta}}}_{(\ell)}^{\star}}{\left(\left(\underbar{\ensuremath{\bm{\theta}}}_{(\ell)}^{\star}\right)^{\dagger}\bm{\Psi}\left(\text{\ensuremath{\underbar{\ensuremath{\bm{a}}}}}_{\mathrm{fix}}\right)\,\underbar{\ensuremath{\bm{\theta}}}_{(\ell)}^{\star}\right)^{2}}+\mathrm{const}\label{eq: MM objective FuCi}
\end{gather}
Conversely, in the IS-based design, the objective function $g_{\mathrm{IS}}(\bm{\theta})$ is convex in $\bm{\theta}$ and can be \emph{minorized} by its first-order Taylor approximation:
\begin{gather}
g_{\mathrm{IS}}(\bm{\theta})=\underbar{\ensuremath{\bm{\theta}}}^{\dagger}\tilde{\bm{\Xi}}\left(\text{\ensuremath{\underbar{\ensuremath{\bm{a}}}}}_{\mathrm{fix}}\right)\underbar{\ensuremath{\bm{\theta}}}\,/\,\left\Vert \text{\ensuremath{\underbar{\ensuremath{\bm{a}}}}}_{\mathrm{fix}}\right\Vert ^{2}\geq\nonumber \\
f_{\mathrm{IS}}(\bm{\theta}|\bm{\theta}_{(\ell)}^{\star})=\Re\left\{ \left(\underbar{\ensuremath{\bm{\theta}}}_{(\ell)}^{\star}\right)^{\dagger}\,\left(\tilde{\bm{\Xi}}\left(\text{\ensuremath{\underbar{\ensuremath{\bm{a}}}}}_{\mathrm{fix}}\right)\,/\left\Vert \text{\ensuremath{\underbar{\ensuremath{\bm{a}}}}}_{\mathrm{fix}}\right\Vert ^{2}\right)\,\underbar{\ensuremath{\bm{\theta}}}\right\} +\mathrm{const}\label{eq: MM objective IS}
\end{gather}
In both Eqs.~\eqref{eq: MM objective FuCi}~and~\eqref{eq: MM objective IS}, \say{$\mathrm{const}$} refers to terms not depending on $\ensuremath{\bm{\theta}}$.
Accordingly, the phase-shift optimization problem in each iteration of the AO can be obtained in both cases as
\begin{gather}
\breve{\mathcal{P}}{}_{\mathrm{FuC,}i}^{(\mathrm{B)}}\,:\,\bm{\theta}_{\mathrm{FuC,i}-(\ell+1)}^{\star}=\underset{|\theta_{m}|=1}{\arg\max}\,f_{\mathrm{FuC},i}(\bm{\theta}|\bm{\theta}_{\mathrm{FuC,}i-(\ell)}^{\star})
\end{gather}
and
\begin{gather}
\breve{\mathcal{P}}{}_{\mathrm{IS}}^{(\mathrm{B)}}\,:\,\bm{\theta}_{\mathrm{IS}-(\ell+1)}^{\star}=\underset{|\theta_{m}|=1,}{\arg\max}\,f_{\mathrm{IS}}(\bm{\theta}|\bm{\theta}_{\mathrm{IS-}(\ell)}^{\star})
\end{gather}
Remarkably, the optimal solution of $\breve{\mathcal{P}}{}_{\mathrm{FuC,}i}^{(\mathrm{B)}}$ is in \emph{closed-form}, namely
\begin{gather}
\angle\text{\ensuremath{\underbar{\ensuremath{\bm{\theta}}}}}_{(\ell+1)}^{\star}=\angle\left(\frac{\bm{\Xi}\left(\text{\ensuremath{\underbar{\ensuremath{\bm{a}}}}}_{\mathrm{fix}}\right)\,\underbar{\ensuremath{\bm{\theta}}}_{(\ell)}^{\star}}{\left(\underbar{\ensuremath{\bm{\theta}}}_{(\ell)}^{\star}\right)^{\dagger}\bm{\Psi}\left(\text{\ensuremath{\underbar{\ensuremath{\bm{a}}}}}_{\mathrm{fix}}\right)\,\underbar{\ensuremath{\bm{\theta}}}_{(\ell)}^{\star}}-\right.\label{eq: RIS optimization (Step 2) fc}\\
\left.\frac{\left(\underbar{\ensuremath{\bm{\theta}}}_{(\ell)}^{\star}\right)^{\dagger}\,\bm{\Xi}\left(\text{\ensuremath{\underbar{\ensuremath{\bm{a}}}}}_{\mathrm{fix}}\right)\,\underbar{\ensuremath{\bm{\theta}}}_{(\ell)}^{\star}}{\left(\left(\underbar{\ensuremath{\bm{\theta}}}_{(\ell)}^{\star}\right)^{\dagger}\bm{\Psi}\left(\text{\ensuremath{\underbar{\ensuremath{\bm{a}}}}}_{\mathrm{fix}}\right)\,\underbar{\ensuremath{\bm{\theta}}}_{(\ell)}^{\star}\right)^{2}}\,\left(\bm{\Psi}\left(\text{\ensuremath{\underbar{\ensuremath{\bm{a}}}}}_{\mathrm{fix}}\right)-\lambda_{max}(\bm{\Psi})\,\bm{I}_{2M}\right)\,\underbar{\ensuremath{\bm{\theta}}}_{(\ell)}^{\star}\right)\nonumber 
\end{gather}
 Conversely, the optimal solution of $\breve{\mathcal{P}}{}_{\mathrm{IS}}^{(\mathrm{B)}}$ is simpler and equal to:
\begin{equation}
\angle\text{\ensuremath{\underbar{\ensuremath{\bm{\theta}}}}}_{(\ell+1)}^{\star}=\angle\left(\tilde{\bm{\Xi}}\left(\text{\ensuremath{\underbar{\ensuremath{\bm{a}}}}}_{\mathrm{fix}}\right)\,\text{\ensuremath{\underbar{\ensuremath{\bm{\theta}}}}}_{(\ell)}^{\star}\right)\label{eq: RIS optimization (Step 2) is}
\end{equation}
For notational brevity, we have dropped the subscripts associated to full-characterization (resp. IS-based design) in Eq.~\eqref{eq: RIS optimization (Step 2) fc} (resp. Eq.~\eqref{eq: RIS optimization (Step 2) is}), namely 
$\bm{\theta}_{\mathrm{\mathrm{FuC},}i-(\ell+1)}^{\star}\rightarrow\bm{\theta}_{(\ell+1)}^{\star}$
and $\bm{\theta}_{\mathrm{FuC},i-(\ell)}^{\star}\rightarrow\bm{\theta}{}_{(\ell)}^{\star}$ (resp. $\bm{\theta}_{\mathrm{IS}-(\ell+1)}^{\star}\rightarrow\bm{\theta}_{(\ell+1)}^{\star}$
and $\bm{\theta}_{\mathrm{IS}-(\ell)}^{\star}\rightarrow\bm{\theta}_{(\ell)}^{\star}$).

\subsection{Summary, Requirements and Complexity Overview} \label{subsec:complexity}

The designed FuC- and IS-based procedures are then summarized in \textbf{Algorithms~\ref{alg:AO-MM-IS}~and~\ref{alg:AO-MM-FC}}.
In both cases the resulting design alternates between the closed-form updates related to fusion rule [\textbf{Step (A)}] and RHS design [\textbf{Step (B)}].

Since the objective function monotonically increases with the iteration number, the procedure is provably convergent in the value of the objective.
Accordingly, the peculiar deflection measure considered (either Eq.~\eqref{eq: deflection_WL} or  Eq.~\eqref{eq: deflection_IS}) is guaranteed to monotonically increase and converge to a \emph{local optimum} by leveraging the structural properties of the AO procedure.
In this work, for simplicity we select the initial point $\bm{\theta}_{(0)}^{\star}$ based on uniformly-generated random phase-shifts.
Nevertheless, other (wiser) initialization strategies are possible.

\begin{algorithm}[htbp]\caption{FuC-based design for Holographic Decision Fusion. Optimization is carried out via AO and MM.}\label{alg:AO-MM-IS}
\begin{mdframed}[backgroundcolor=lblue]
\begin{algorithmic}[1]
\STATE Construct an initial $\bm{\theta}_{(0)}^{\star}$ and set $\ell=0$;
\REPEAT
\STATE Fix $\bm{\theta}_{(\ell)}^{\star}$ and update $\underbar{\ensuremath{\bm{a}}}$ according to Eq.~\eqref{eq: WL fusion beamformer Step A - FC};
\STATE Fix $\underbar{\ensuremath{\bm{a}}}$ and update $\bm{\theta}_{(\ell+1)}^{\star}$ via Eq.~\eqref{eq: RIS optimization (Step 2) fc};
\STATE Set $\ell \xleftarrow{} \ell+1$;
\UNTIL convergence 
\end{algorithmic}
\end{mdframed}
\end{algorithm}

\begin{algorithm}[htbp]
\caption{IS-based design for Holographic Decision Fusion. Optimization is carried out via AO and MM.}
\label{alg:AO-MM-FC}
\begin{mdframed}[backgroundcolor=lred]
\begin{algorithmic}[1]
\STATE Construct an initial $\bm{\theta}_{(0)}^{\star}$ and set $\ell=0$;
\REPEAT
\STATE Fix $\bm{\theta}_{(\ell)}^{\star}$ and update $\underbar{\ensuremath{\bm{a}}}$ according to Eq.~\eqref{eq: WL fusion beamformer Step A - IS};
\STATE Fix $\underbar{\ensuremath{\bm{a}}}$ and update $\bm{\theta}_{(\ell+1)}^{\star}$ via Eq.~\eqref{eq: RIS optimization (Step 2) is};
\STATE Set $\ell \xleftarrow{} \ell+1$;
\UNTIL convergence 
\end{algorithmic}
\end{mdframed}
\end{algorithm}

The \emph{computational complexity} of the proposed AO-based joint design is thus $\mathcal{O}(N_{\mathrm{iter}}\,(C_{\mathrm{fus}}+C_{\mathrm{rhs}} ) + C_{\mathrm{init}})$, where $N_{\mathrm{iter}}$ denotes the number of \emph{outer} iterations of the AO procedure (i.e. each consisting of a \textbf{Step (A)} and a \textbf{Step (B)}).
Conversely, $C_{\mathrm{fus}}$ and $C_{\mathrm{rhs}}$ denote the cost of the WL vector fusion and RHS matrix update steps, respectively.
Finally $C_{\mathrm{init}}$ denotes the cost of static operations which can be performed outside the AO loop.
The cost of each is reported in Tab.~\ref{tab: Complexity comparison} for \emph{both} design approaches considered. 

Regarding \textbf{Step (A)}, the complexity for the IS-based design is mainly dominated by the multiplication between $\bm{G}$ and the $M\times 1$ vector $\bm{\Theta}_{\mathrm{fix}}\bm{H} \bm{D}_{\alpha}\bm{1}_{K}$ (assuming the term $\bm{H} \bm{D}_{\alpha}\bm{1}_{K}$ is pre-computed at the beginning of the procedure).
Conversely, for the FuC-based design, the complexity is mainly given by the matrix products needed to obtain $\bm{H}^{e}(\bm{\Theta})$ and the augmented covariance $\mathrm{Cov}_{\bm{\Theta}_{\mathrm{fix}}}(\text{\ensuremath{\underbar{\ensuremath{\bm{y}}}}}|\mathcal{H}_{i})$ (cf. Eq.~\eqref{eq: augmented covariance rx signal vector}), as well as to invert the latter matrix.

Regarding \textbf{Step (B)}, the complexity of the IS-based design primarily arises from performing matrix-vector products, specifically to compute $\bm{N}^{\dagger}\text{\ensuremath{\underbar{\ensuremath{\bm{a}}}}}_{\mathrm{fix}}$ and 
$\tilde{\bm{\Xi}}\left(\text{\ensuremath{\underbar{\ensuremath{\bm{a}}}}}_{\mathrm{fix}}\right)\,\text{\ensuremath{\underbar{\ensuremath{\bm{\theta}}}}}_{(\ell)}^{\star}$, and constructing the outer product matrix $\tilde{\bm{\Xi}}\left(\text{\ensuremath{\underbar{\ensuremath{\bm{a}}}}}_{\mathrm{fix}}\right)$.
Conversely, the complexity of the FuC-based design is dominated by the computation of $\lambda_{max}(\bm{\Psi})$, evaluating $\bm{\Psi}\left(\text{\ensuremath{\underbar{\ensuremath{\bm{a}}}}}_{\mathrm{fix}}\right)$ and determining $\bm{\Delta}_{0}$.

Furthermore, it is worth noticing that the reported complexity of Steps (A-B) for IS-based design \emph{does not take into account terms which are static} (last column of Tab.~\ref{tab: Complexity comparison}) and can be evaluated outside the AO loop, i.e. computation of $\bm{H}\,\bm{D}_{\alpha}\bm{1}_{K}$ for the fusion step while the computation of $\bm{N}_{r} = \bm{G}\,\mathrm{diag}(\bm{H}\,\bm{D}_{\alpha}\,\bm{\rho}_{10})$ for the RHS update step.
Once the design procedure has been carried out, it is worth noticing that given the WL vector assumption made in this work (cf. Eq.~\eqref{eq: WL fusion statistic}), the complexity of the DD task at the FC is $\mathcal{O}(N)$ (since the RHS carries out analog processing) as opposed to an LLR-based implementation that would require $\mathcal{O}(N\,2^{K})$ (cf. Eq.~\eqref{eq:LLR_RIS}).

Last, by comparing the \emph{required parameters} for implementation of the FuC and IS design strategies (see Tab.~\ref{tab: knowledge parameters}), it is apparent that the latter (by design) does not require sensing parameters, as opposed to FuC.
Still, FuC only requires sensing statistical characterization up to the second order (mean vector and covariance).
This contrasts with the usual fusion rule implementation via the LLR (requiring the joint pmfs $\Pr(\bm{x}|\mathcal{H}_{1})$ and $\Pr(\bm{x}|\mathcal{H}_{0})$).

\renewcommand{\arraystretch}{1.7}
\begin{table*}
\begin{centering}
\medskip{}
\par\end{centering}
\centering{}\caption{Summary of the computational complexity needed for joint
FuC- and IS-based design strategies. For each strategy, the complexity of the fusion ($C_{\mathrm{fus}}$) and RHS design steps ($C_{\mathrm{rhs}}$) involved in the AO are underlined, together with the complexity needed outside the AO loop ($C_{\mathrm{init}}$). $K$ denotes the number of sensors in the WSN; $M$ denotes the number of RHS reflecting elements, whereas $N$ represents
the number of antennas feeds. \label{tab: Complexity comparison}}
\begin{tabular}{cccc}
\hline 
\textbf{Design Principle} & \textbf{Complexity AO Step A ($C_{\mathrm{fus}}$)} & \textbf{Complexity AO Step B ($C_{\mathrm{rhs}}$)} & \textbf{Init Complexity ($C_{\mathrm{init}}$)}\tabularnewline
\hline 
\rowcolor[HTML]{edf7ff}FuC & $\ensuremath{\mathcal{O}(M^2K+NMK+K^2N+KN^2+N^3)}$  & $\ensuremath{\ensuremath{\mathcal{O}(M^{3}+MN+K^{2}M+M^{2}K+KN)}}$  & -- \tabularnewline
\rowcolor[HTML]{f8e8e8}IS & 
$\mathcal{O}(MN)$  & $\mathcal{O}(M^{2}+MN)$ & $\mathcal{O}(NK+MK+NM)$\tabularnewline
\hline 
\end{tabular}
\end{table*}
\renewcommand{\arraystretch}{1}

\renewcommand{\arraystretch}{1.7}
\begin{table}
\begin{centering}
\medskip{}
\par\end{centering}
\centering{}\caption{Overview of the channel and sensing parameters required for FuC and IS design strategies. For reference, the system knowledge necessary for implementing the LLR combined with FuC/IS-designed RHS matrix is also included (first row).\label{tab: knowledge parameters}}
\begin{tabular}{cccc}
\hline 
\textbf{Design Principle} & \textbf{Knowledge Requirements}\tabularnewline
\hline 
\rowcolor[HTML]{DCDCDC}LLR + FuC/IS-RHS & $\Pr(\bm{x}|\mathcal{H}_{1})$, $\Pr(\bm{x}|\mathcal{H}_{0})$, $\bm{G}$, $\bm{H}$, $\bm{D}_{\alpha}$, $\sigma_{w}^{2}$ \tabularnewline
\rowcolor[HTML]{edf7ff}FuC-1 & $\bm{\rho}_{1}$, $\bm{\rho}_{0}$, $\mathrm{Cov}\left(\bm{x}|\mathcal{H}_{1}\right)$, $\bm{G}$, $\bm{H}$, $\bm{D}_{\alpha}$, $\sigma_{w}^{2}$  \tabularnewline
\rowcolor[HTML]{edf7ff}FuC-0 & $\bm{\rho}_{1}$, $\bm{\rho}_{0}$, $\mathrm{Cov}\left(\bm{x}|\mathcal{H}_{0}\right)$, $\bm{G}$, $\bm{H}$, $\bm{D}_{\alpha}$, $\sigma_{w}^{2}$  \tabularnewline
\rowcolor[HTML]{f8e8e8}IS & $\bm{G}$, $\bm{H}$, $\bm{D}_{\alpha}$  \tabularnewline
\hline 
\end{tabular}
\end{table}
\renewcommand{\arraystretch}{1}

\section{Simulation Results}
\label{Sim_res}
\subsection{Considered Setup}
When not otherwise specified, we consider a WSN made of $K=10$ sensors whose local decisions on the phenomenon of interest are conditionally independent and identically distributed (i.i.d.), i.e.,
$\Pr(\bm{x}|\mathcal{H}_{i})=\prod_{k=1}^{K}\Pr(x_{k}|\mathcal{H}_{i})$, where $(P_{D,k},P_{F,k})\triangleq(P_{D},P_{F})=(0.5,0.05)$, $k\in\mathcal{K}$.
Similar values have been also used in related studies on DD~\cite{Chen2004, ciuonzo2015}.
When the i.i.d. assumption holds, the conditional mean vector and covariance matrix in Eq.~\eqref{eq:2nd_order_char} have the simplified expressions $\bm{\rho}_{i}=P_{i}\,\bm{1}_{K}$ and $\mathrm{Cov}\left(\bm{x}|\mathcal{H}_{i}\right)=4P_{i}(1-P_{i})\bm{I}_{K}$, respectively.

For simplicity, we express all distances/spacings in multiples/fraction of the wavelength $\lambda$.
The locations of the WSN, RHS and receive feeds are as follows.
The sensors are uniformly distributed at random in the box $[0,40]\times[0,40]\times[0,3] \lambda$ (corresponding to x-, y- and z-axes).
The RHS is a square planar surface with spacing $\lambda/3$.
Furthermore, the RHS has its geometrical center located at $\bar{\bm{p}}^{\mathrm{rhs}}=[70, 20, 10]\, \lambda$ and it is displaced parallel to the y-axis.
Conversely, the $N$ (external) receive feeds are arranged in a linear configuration with spacing $\lambda/2$.
The feeds are centered around the coordinate $[68, 18, 10]\,\lambda$ in parallel to the x-axis.
For simplicity, we assume that the element side lengths coincide with the corresponding element spacings, i.e. $\Delta_{\mathrm{h}}^{\mathrm{rhs}}=\Delta_{\mathrm{v}}^{\mathrm{rhs}}=\Delta^{\mathrm{rhs}}=\lambda/3$
and $\Delta_{\mathrm{h}}^{\mathrm{fc}}=\Delta_{\mathrm{v}}^{\mathrm{fc}}=\Delta^{\mathrm{fc}}=\lambda/2$, respectively.

In such a configuration, the Fraunhofer distance for the planar square RHS composed of $M$ elements, as well as for a 1-D linear array of $N$ feeds is given by~\cite{interdonato2024approaching}:
\begin{equation}
d_{\mathrm{fr}}\triangleq\frac{2}{\lambda}\max\left\{ (\sqrt{M}\Delta^{\mathrm{rhs}})^{2},(N\Delta^{\mathrm{fc}})^{2}\right\} 
\end{equation}
For instance, considering an RHS with $M=100$, along with $N=2$ receive feeds, we obtain $d_{\mathrm{fr}} \approx 22\lambda$.
If the RHS is reduced to $M=25$ elements, this distance decreases to $d_{\mathrm{fr}} \approx 5.5 \lambda$.
By contrast, the geometric center separation between the RHS and the array of feeds is $ \approx 2.8 \lambda$.
Since this distance is well below the Fraunhofer threshold, wave propagation occurs in the near-field regime, justifying the channel model expressed in Eq.~\eqref{eq:channel_model}.

For simplicity, the sensors are assumed to have equal transmit energy, namely $\alpha_k=1$.
Also, their path loss attenuation at the reference distance of $d_0=1\,\lambda $ is set to $\mu = - 30\,\mathrm{dB}$. Furthermore, the path loss exponent is set to $\nu=2$.  
The parameter $b_k$ is defined as $b_{k}\triangleq\sqrt{\frac{\kappa_{k}}{1+\kappa_{k}}}$, where $\kappa_{k}$ denotes the Rician factor for the $k$th sensor, randomly drawn from the range $(3,5)\,\mathrm{dB}$.

The value $q$ in Eq.~\eqref{eq: general directivity pattern} is chosen to enforce a $\cos^3(\cdot)$ gain profile: in such a case the maximum gain equals $8$ ($\approx 9.03\, \mathrm{dBi}$).
For simplicity, the reflection efficiency of each RHS element is set to $\eta=1$.
Finally, the noise variance $\sigma_{w}^{2}$ is set to $-50$ dBm in Eq.~\eqref{eq: signal_model}.

\subsection{Baselines and Upper bound}

The WSN system performance is evaluated in terms of the global probabilities of \textit{false alarm} $P_{F_{0}}\triangleq\Pr\{\Lambda>\gamma|H_{0}\}$ and  \textit{detection} $P_{D_{0}}\triangleq\Pr\{\Lambda>\gamma|H_{1}\}$.
In the following analysis, the \emph{observation bound} is also reported for completeness (curve \say{Obs. bound} in the following figures). The latter represents the performance of the optimal decision fusion rule in an ideal channel condition, given by (in the i.i.d. case)
\begin{eqnarray}
P_{D_{0}}^{\text{ob}} & = & \sum_{i=\nu}^{K}\binom{K}{i}(P_{D})^{i}\,(1-P_{D})^{K-i}\label{eq: obs_bound}\\
P_{F_{0}}^{\text{ob}} & = & \sum_{i=\nu}^{K}\binom{K}{i}(P_{F})^{i}\,(1-P_{F})^{K-i}\nonumber 
\end{eqnarray}
where $\nu\in\{0,\ldots K\}$ is a discrete threshold.
The above bound represents a relevant benchmark to assess both ($i$) the detection degradation due to the interfering distributed MIMO channel and ($ii$) the corresponding benefit arising from the RHS adoption.

\begin{figure*}[ht]
    \centering
    
    \begin{subfigure}[b]{0.45\textwidth}
       \centering{}\includegraphics[width=0.95\columnwidth]{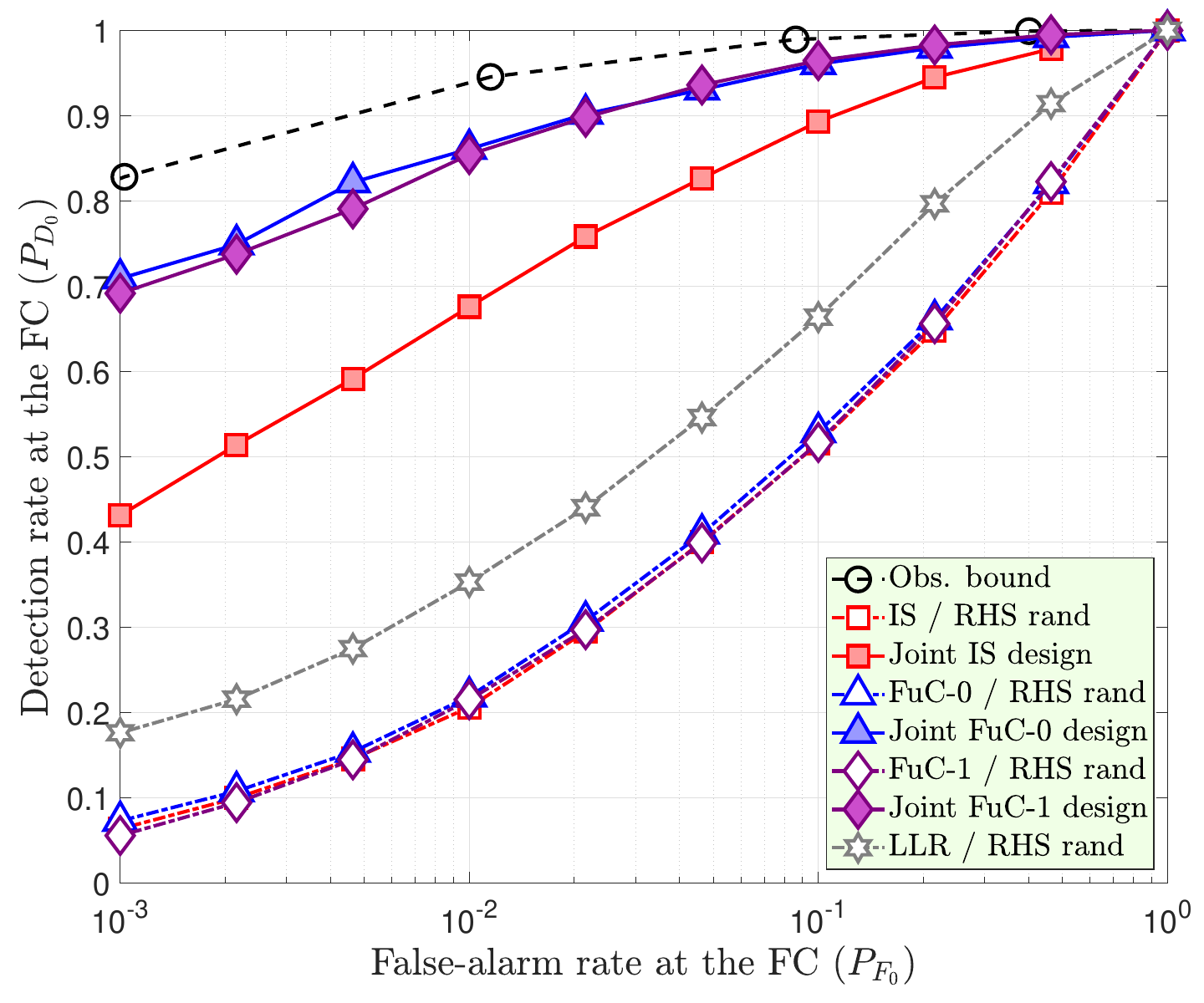}\caption{Comparison of joint design strategies and fusion rule counterparts with random RHS configurations. For reference, also the LLR rule with random RHS configuration is reported. \label{fig:ROC_RIS_design}}
    \end{subfigure}
    \hfill
    \begin{subfigure}[b]{0.45\textwidth}
       \centering{}\includegraphics[width=0.95\columnwidth]{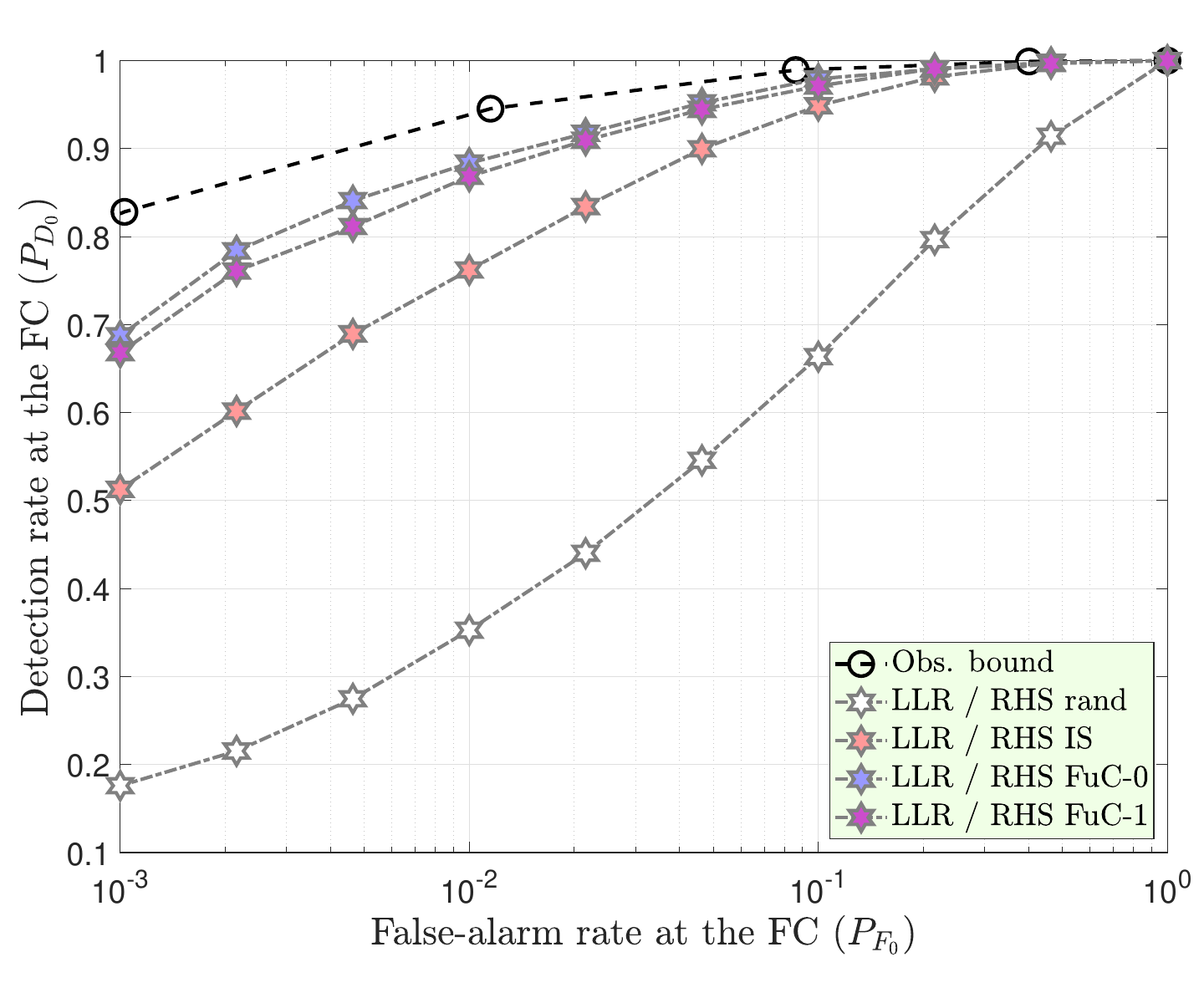}\caption{Comparison of the LLR rule performance with different RHS design strategies. As a baseline, the LLR rule with random RHS configuration is also reported. \label{fig:ROC_RIS_LLRwdesign}}
    \end{subfigure}
    
    \caption{Assessing the benefits of joint design in holographic DF via ROCs ($P_{D_{0}}$ vs $P_{F_{0}}$). WSN with $K=10$ sensors, $(P_{D,k},P_{F,k})=(0.5,0.05)$, $k\in\mathcal{K}$. Holographic DF is implemented with an RHS made of $M=64$ elements and $N=1$ receive feed; noise variance is set to $\sigma_{w}^{2}=-50$ dBm.}
    \label{fig:PD0_vs_M_combined}
\end{figure*}

\subsection{Assessing Benefits of Joint Design in Holographic DF}

Figure~\ref{fig:ROC_RIS_design} illustrates the Receiver Operating Characteristic (ROC), namely $P_{D_{0}}$ vs $P_{F_{0}}$, for the considered rules in a WSN implementing holographic DF via a RHS composed of $M=64$ elements and a single feed ($N=1$).
Performance analysis reveals that the three proposed joint design strategies achieve \emph{notable performance gains} compared to scenarios where the fusion rule is designed without optimizing the RHS~\cite{Ciuonzo2012,ciuonzo2015}.
These correspond to FuC-1, FuC-0 and IS design strategies when only the fusion step is carried out while considering the RHS shifts as set uniformly at random.
For reference, the performance of the log-likelihood ratio (LLR) fusion statistic under non-optimized RHS conditions (denoted with \say{LLR / RHS rand}) is also presented, implemented via Eq.~\eqref{eq:LLR_RIS}. 
The latter \emph{represents the best performance that can be attained when the RHS is not designed}.
These results underscore the key role of joint design in achieving superior global detection capabilities.

As anticipated, the design strategies based on full characterization (namely FuC-1 and FuC-0) consistently surpass the approach grounded in ideal sensing (IS). 
This superiority is attributed to the fact that the IS method overlooks the statistical properties of the sensing process, treating it as idealized (cf. Eq.~\eqref{eq: 2nd order char IS}). 
Despite this, the joint design approach \emph{always demonstrates a compelling performance boost}. For instance, relative to the LLR with non-optimized RHS, detection rate improvements (when the false-alarm rate is set to $P_{F_{0}}=0.01$) of approximately $30\%$, and $55\%$ are observed for IS and FuC, respectively.

Given that the proposed joint design solutions impose a constraint on the fusion rule, limiting it to a WL statistic (cf. Eq.~\eqref{eq: WL fusion statistic}), it is essential to examine the impact of the proposed RHS design strategies when applying the LLR statistic defined in Eq.~\eqref{eq:LLR_RIS}. 
To this end, in Fig.~\ref{fig:ROC_RIS_LLRwdesign} the ROC of the LLR is compared under two conditions: with non-optimized RHS (where phase shifts are randomly selected within $[0,2\pi)$) and with optimized RHS.
The findings highlight that, although the proposed joint design strategies are not explicitly tailored for the LLR statistic, they still offer significant performance gains in this context. Interestingly, the same performance hierarchy observed for the WL fusion statistic (FuC-0 $>$ FuC-1 $>$ IS) is mirrored when the LLR statistic is used, as seen in Fig.~\ref{fig:ROC_RIS_design}.

\subsection{Performance Trends vs. Number of RHS Elements}
Then, in Fig.~\ref{fig:PD0_vs_M_combined} we report the $P_{D_{0}}$ (for a fixed false-alarm rate, set to $P_{F_{0}}=0.01$) versus the number of RHS elements ($M$) to investigate the improvement achievable considering a larger RHS in the proposed holographic DF setup.

The proposed system is compared against a fully-digital MIMO architecture at the FC~\cite{Ciuonzo2012,ciuonzo2015} equipped with $N_{dig}=100$ antennas and RF chains. 
For the sake of a fair comparison, the digital architecture has the $N_{dig}$ antennas arranged in a square planar array fashion with $\lambda/2$ spacing and whose geometric center is the same as $\bar{\bm{p}}^{\mathrm{rhs}}$ and same planar alignment as the RHS.
For the fully-digital MIMO FC, three fusion rules are evaluated, corresponding to the same design principles as the proposed system (cf. Sec.~\ref{subsec: fusion_rule_design}). 
The resulting detection performance is depicted using yellow-filled markers.

First, by looking at the trend with $M$, all the three design strategies improve the detection performance as the number of elements increases.
Of course, the asymptotic performance depends on the peculiar design strategy, with each holographic design reaching approximately the performance of its fully-digital counterpart as $M$ increases.
Clearly, the IS design in the fully-digital case is again limited by the idealized sensing assumption.
For instance, when $M=144$ and $N=1$ is employed, the detection rate of the proposed holographic DF is only $\approx 2\%$ lower when using FuC-0 and IS strategies, and $\approx 4\%$ when using FuC-1 strategy. 
No appreciable performance difference is observed for a large RHS when increasing the number of receive feeds considered from $N=1$ to $N=2$.
Conversely, for a smaller number of RHS elements (i.e. $M\in (25, 49)$), there is an appreciable improvement in using $N=2$ feeds.

It is worth noticing that the \emph{holographic DF power consumption} is expressed as  
\begin{gather}
\epsilon_{\mathrm{holo}}=\underbrace{\sum_{k=1}^{K}\alpha_{k}^{2}+K\cdot\epsilon_{\mathrm{tx}}^{\mathrm{sen}}\,}_{\epsilon_{\mathrm{tx}}}+\underbrace{M\cdot\epsilon_{\mathrm{rhs}}+N\cdot\epsilon_{\mathrm{rx}}^{\mathrm{\mathrm{feed}}}}_{\epsilon_{\mathrm{rx},\mathrm{holo}}}+\epsilon_{0}
\end{gather}
where $\epsilon_{\mathrm{tx}}^{\mathrm{sen}}$, $\epsilon_{\mathrm{rhs}}$, and $\epsilon_{\mathrm{rx}}^{\mathrm{feed}}$ denote the power consumption of each sensor’s antenna, reflecting element, and receive feed, respectively, and $\epsilon_{0}$ accounts for static system power consumption.
Conversely, the \emph{power expenditure of the fully-digital DF} is given by
\begin{eqnarray}
\epsilon_{\mathrm{dig}} & =\underbrace{\sum_{k=1}^{K}\alpha_{k}^{2}+K\cdot\epsilon_{\mathrm{tx}}^{\mathrm{sen}}}_{\epsilon_{\mathrm{tx}}}+ & \underbrace{N_{\mathrm{dig}}\cdot\epsilon_{\mathrm{rx}}^{\mathrm{feed}}}_{\epsilon_{\mathrm{rx},\mathrm{dig}}}\,+\,\epsilon_{0}
\end{eqnarray}
Both configurations share identical power consumption during the reporting phase (denoted with $\epsilon_{\mathrm{tx}}$), but the receive phase consumption differs ($\epsilon_{\mathrm{rx,holo}} \neq \epsilon_{\mathrm{rx,dig}}$).
For instance, considering an RHS with $M=144$, $N=1$, and typical values such as $\epsilon_{\mathrm{rf}} \approx 10\epsilon_{\mathrm{rhs}}$~\cite{zappone2022energy}, the ratio of the receive-phase power expenditures is approximately $\epsilon_{\mathrm{rx,dig}} / \epsilon_{\mathrm{rx,holo}} \approx \frac{(100 \cdot 10)\epsilon_{\mathrm{rhs}}}{(144 + 10)\epsilon_{\mathrm{rhs}}} \approx 6.5$.
Thus, the proposed holographic DF system \emph{achieves comparable detection performance to the fully-digital case while being approximately $6.5\times$ more energy-efficient at the FC side}.

\begin{figure*}[ht]
    \centering
    
    \begin{subfigure}[b]{0.45\textwidth}
        \centering
        \includegraphics[width=\textwidth]{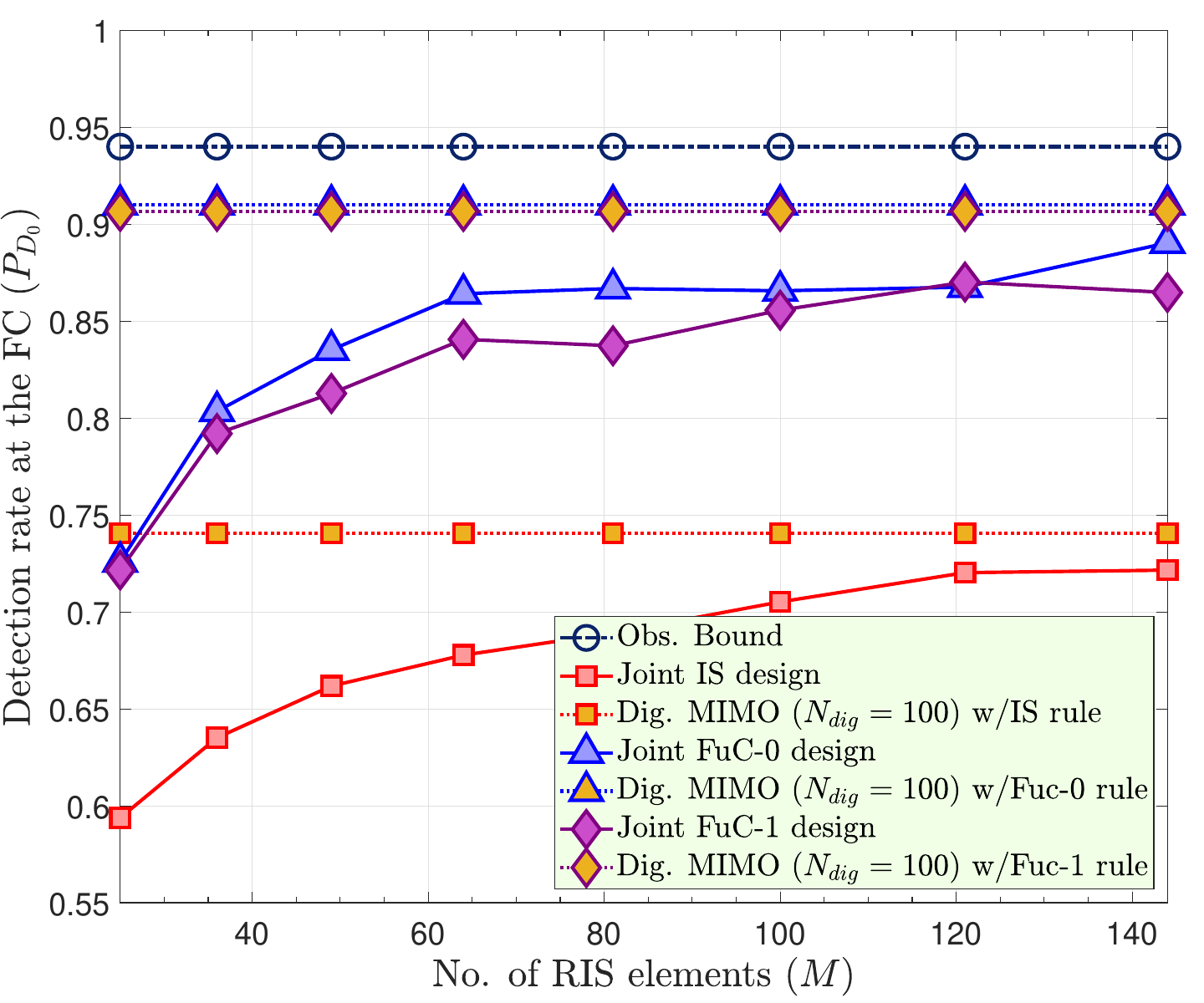}
        \caption{$N=1$ receive feed is considered.}
        \label{fig:PD0_vs_M_N1}
    \end{subfigure}
    \hfill
    \begin{subfigure}[b]{0.45\textwidth}
        \centering
        \includegraphics[width=\textwidth]{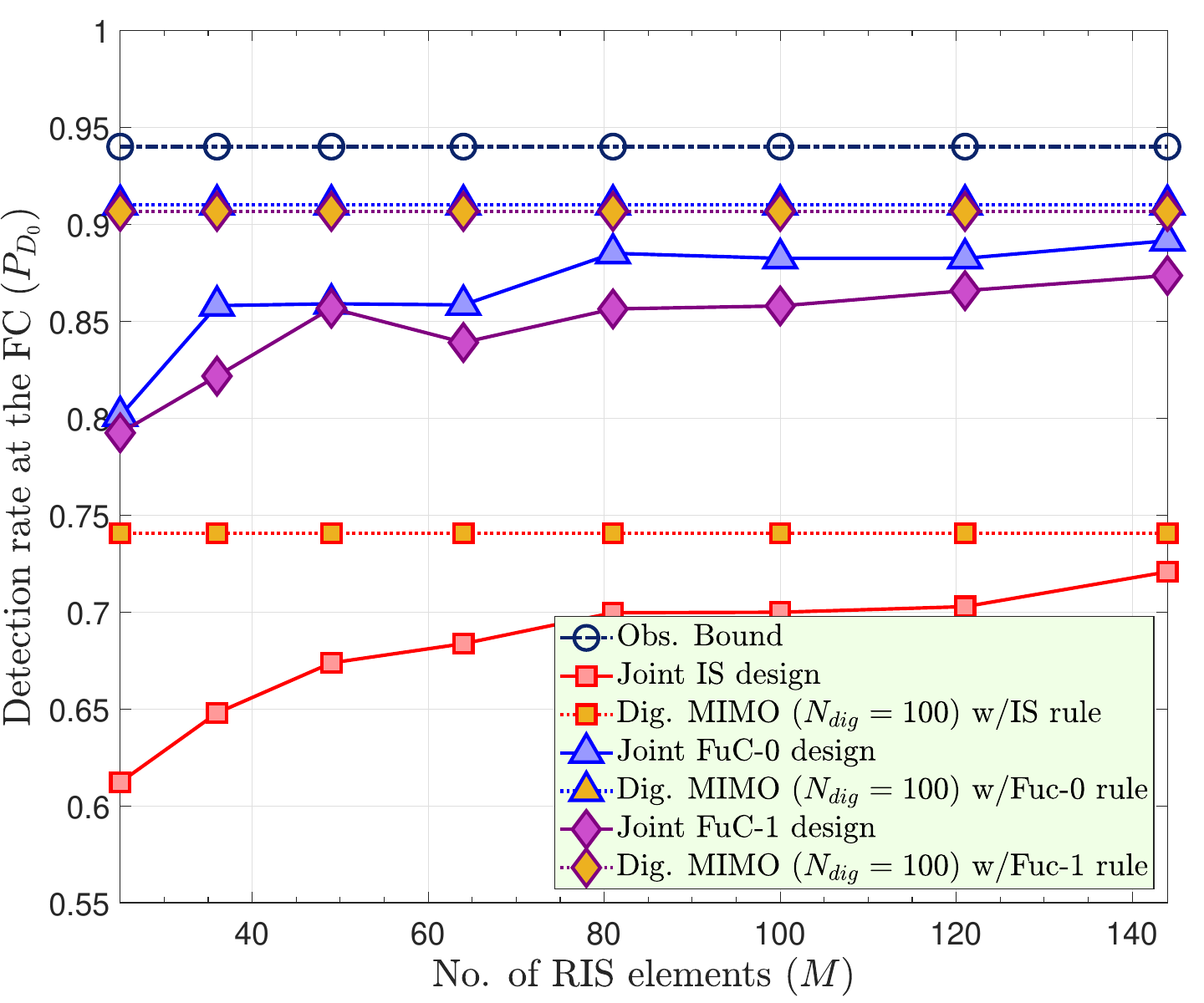}
        \caption{$N=2$ receive feeds are considered.}
        \label{fig:PD0_vs_M_N2}
    \end{subfigure}
    
    \caption{$P_{D_{0}}$ vs number of RHS elements $M$ (with $P_{F_{0}}=0.01$) for two different feed scenarios ($N\in{1,2}$). WSN with $K=10$ sensors, with sensing performance $(P_{D,k},P_{F,k})=(0.5,0.05)$, $k\in\mathcal{K}$. noise variance is set to $\sigma_{w}^{2}=-50$ dBm.}
    \label{fig:PD0_vs_M_combined}
\end{figure*}

\subsection{Holographic DF Performance vs. Number of Fused Sensors}
\begin{figure}
\centering{}\includegraphics[width=0.95\columnwidth]{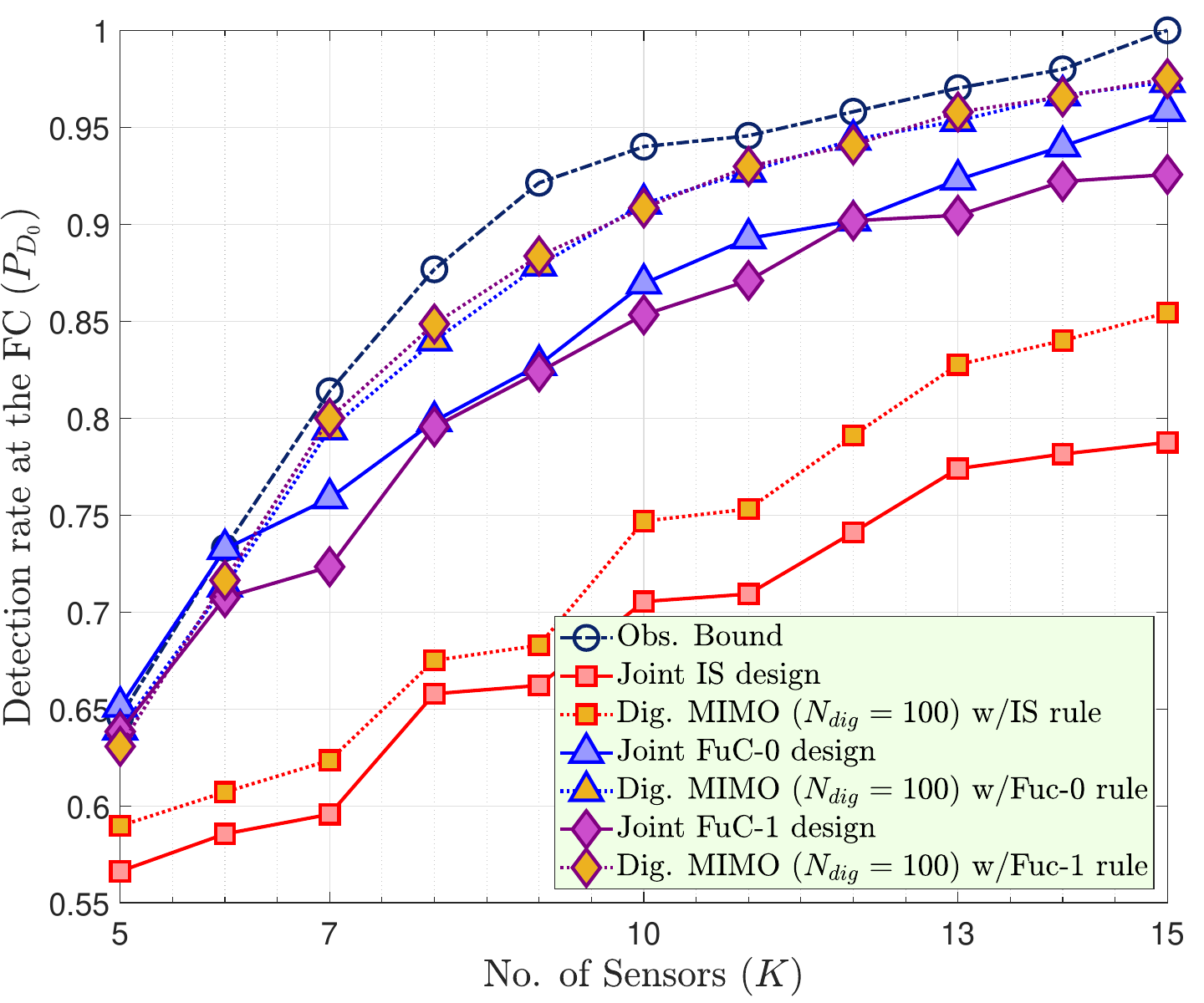}\caption{$P_{D_{0}}$ vs number of sensors $K$ (with $P_{F_{0}}=0.01$) of the considered rule/RHS configurations. Holographic DF is implemented with an RHS made of $M=100$ elements and $N=1$ receive feed; Fully-digital MIMO architecture is simulated with $N_{dig}=100$ antennas. Noise variance is set to $\sigma_{w}^{2}=-50$ dBm. \label{fig:Pd0_vs_K}
}
\end{figure}

Figure~\ref{fig:Pd0_vs_K} evaluates the ability of the proposed holographic DF approach to leverage an increasing number of sensors ($K$), to enhance global detection performance. The figure presents $P_{D_{0}}$ (at a fixed false-alarm rate, $P_{F_{0}} = 0.01$) as the WSN size varies from $K=5$ to $K=15$ sensors.
The holographic DF system, with $M=100$ RHS elements and $N=1$ receive feed, is compared against a fully-digital MIMO counterpart featuring $N_{dig}=100$ antennas. 
It is worth noticing that in this case also the observation bound grows as $K$ increases (cf. Eq.~\eqref{eq: obs_bound}).

Fusion rules using the fully-digital approach effectively exploit the larger number of sensors~\cite{ciuonzo2015}, with FuC-0 and FuC-1 demonstrating minimal loss relative to the observation bound, consistent with prior findings in large-array scenarios~\cite{ciuonzo2015}. 
Meanwhile, the proposed joint design solutions achieve performance close to their fully-digital counterparts, utilizing significantly fewer computational resources and offering greater energy efficiency (i.e., a single receive feed versus $N_{dig}=100$ in the fully-digital case). 
The holographic DF approach also benefits from improved sensing diversity as $K$ increases. For instance, the FuC-1 design achieves a detection rate improvement from approximately $87\%$ to $95\%$ when moving from $K=10$ to $K=15$ sensors.  
Equally important, the performance gap remains consistent across the entire range of WSN sizes considered, including the upper end at $K \in \{14, 15\}$.

\subsection{Performance Loss with Discrete Phases}
\begin{figure*}
\begin{centering}
\includegraphics[width=0.90\paperwidth]{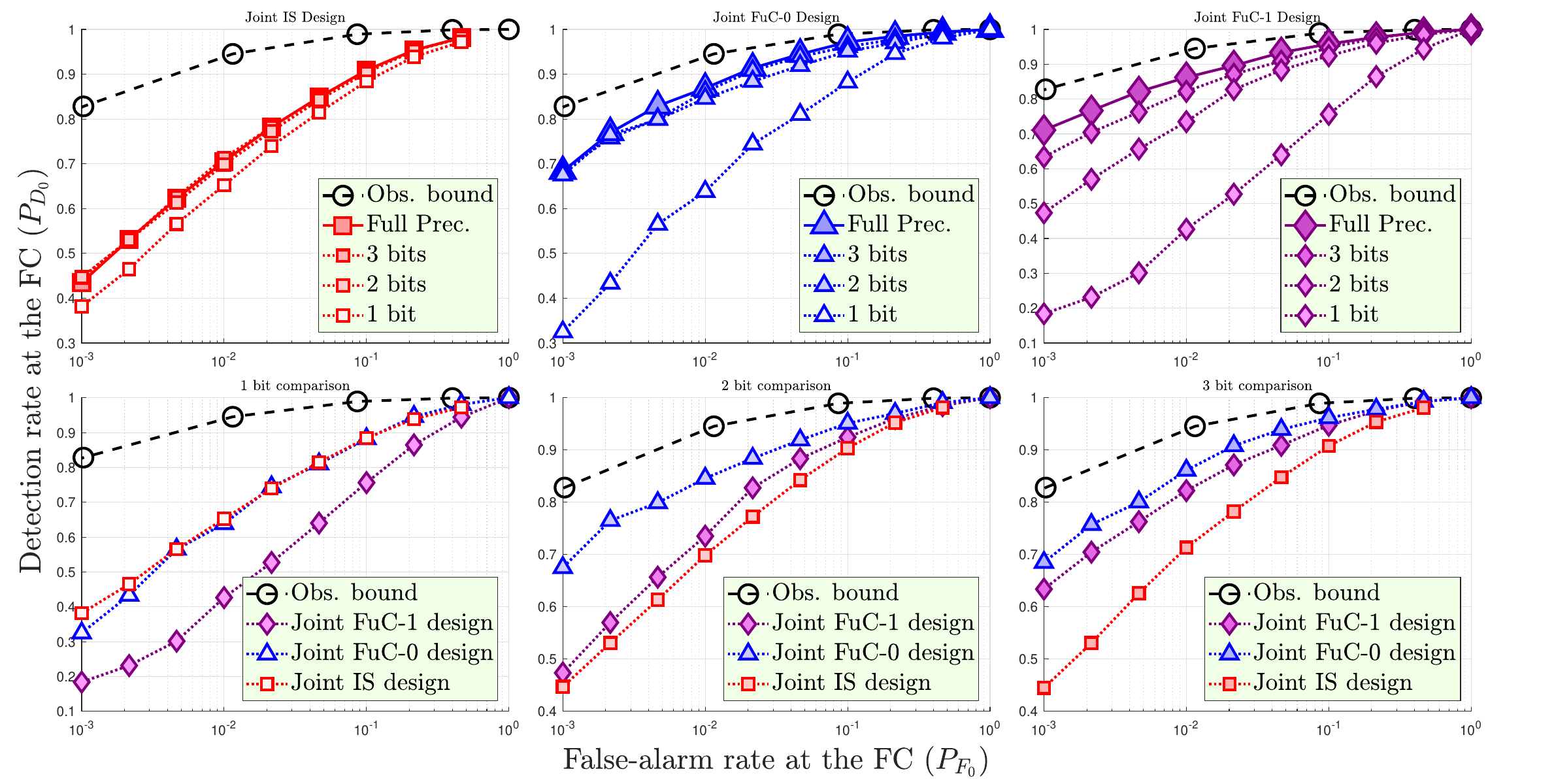}
\caption{$P_{D_{0}}$ vs $P_{F_{0}}$ of the proposed joint design strategies (Fuc-1, FuC-0 and IS) with different resolution of quantized RHS shifts. WSN with $K=10$ sensors, $(P_{D,k},P_{F,k})=(0.5,0.05)$, $k\in\mathcal{K}$.Holographic DF is implemented with an RHS made of $M=100$ elements and $N=1$ receive feed; noise variance is set to $\sigma_{w}^{2}=-50$ dBm.
The first row compares each design strategy at different quantization resolutions to its full-precision counterpart, while the second row contrasts the detection performance of the three strategies at a fixed resolution.
\label{fig:ROC_quantization}}
\end{centering}
\end{figure*}

Figure~\ref{fig:ROC_quantization} illustrates the ROC for the proposed joint design strategies (IS, FuC-0, and FuC-1) under quantization of the RHS design outcomes ($\bm{\theta}$) generated by Algos.~\eqref{alg:AO-MM-IS} and~\eqref{alg:AO-MM-FC}. 
The number of quantization bits used to encode the phase values of the RHS coefficients is varied, and the resulting detection performance is compared against their full-precision counterparts (i.e., no quantization applied). This analysis evaluates the impact of finite resolution in the codebook for the RHS configuration.
The results reveal that one-bit quantization introduces noticeable detection performance degradation, with the extent of the loss varying by design strategy (top row). 
FuC-1 and FuC-0 are more sensitive to quantization, whereas the IS-based design exhibits strong robustness, achieving near-full-precision performance with just two bits. 
Remarkably, IS outperforms FuC-1 with only one-bit quantization and performs similarly to FuC-0 under the same conditions (bottom row). FuC-0, while more affected by quantization than IS, still experiences less degradation compared to FuC-1.
Overall, \emph{three-bit quantization is sufficient to effectively encode the RHS phases across all three design strategies with minimal performance loss}.

\section{Conclusions and Future Directions}
\label{Conclusions}
This work is a first attempt in the use of RHS for aiding channel-aware decision fusion in a distributed MIMO setup via the design of a holographic DF approach.
Recognizing the computational challenges and the absence of closed-form solutions for the LLR fusion rule, we proposed three novel strategies (FuC-0, FuC-1 and IS) for the joint design of RHS configurations (for the analog processing part) and fusion rules (for the digital processing part).
These strategies are anchored in two core design principles.
The first strategy utilizes the second-order statistical characterization of the received data (FuC-0 and FuC-1), while the second adopts a sensor-agnostic approach, jointly optimizing RHS and fusion rules under the assumption of ideal WSN sensing operations (IS). Given the inherent non-convexity of the resulting optimization problems, we addressed them using AO and MM frameworks. This approach yielded a straightforward ping-pong optimization procedure with closed-form updates.
The complexity and system knowledge requirements of the proposed design strategies have been thoroughly evaluated and documented for completeness.
Simulation results highlight the effectiveness of the proposed design solutions, achieving similar detection performance as a fully-digital FC equipped with a large antenna array (e.g., $N_{dig}=100$) while offering substantial energy efficiency gains (e.g. $6.5\times$ reduction in the power consumption at the FC side).
Furthermore, the proposed methods showcase the capacity to harness larger sensor networks ($K$) and exploit larger RHS surfaces (increasing $M$), highlighting the scalability of the approach.
A noteworthy observation is that, despite the continuous-valued outputs stemming from the AO+MM optimization, practical implementation on RHS hardware requires only a small number of quantization bits.
Remarkably, it has been shown that just $3$-bit quantization suffices to achieve negligible performance loss compared to their full-precision counterparts.
Conversely, more stringent quantizations may invert the relative performance of the considered design strategies.

Looking ahead, \emph{future research} directions are poised to further expand the potential of holographic architectures.
This includes the exploration of dynamic metasurfaces or stacked intelligent metasurfaces and the direct design of RHS configurations under practical constraints (e.g., discrete phase shift alphabets).
A key focus will be also adapting to \emph{varying levels of CSI availability}. We will address cases of imperfect CSI by developing specialized channel estimation protocols tailored for holographic DF. For partial CSI scenarios, we will explore two-timescale protocols that leverage (efficient) estimation of the sole sensor-RHS-FC composite channel. Last, we will consider setups where both the FC and RHS design rely on long-term channel statistics.

\section{Acknowledgements}
The research leading to these results has received funding from Project ``Garden'', CUP H53D23000480001 funded by EU in NextGeneration EU plan, Mission 4 Component 1, through the Italian ``Bando Prin 2022 - D.D. 104 del 02-02-2022`` by MUR.
This manuscript reflects only the authors’ views and opinions and the Ministry cannot be considered responsible for them.

The work of A. Zappone has been funded by the European Union - NextGenerationEU under the project NRRP RESTART, RESearch and innovation on future Telecommunications systems and networks, to make Italy more smART PE\_00000001 - Cascade Call SMART project, with CUP E63C22002040007. 

The work of M. Di Renzo was supported in part by the European Union through the Horizon Europe project COVER under grant agreement number 101086228, the Horizon Europe project UNITE under grant agreement number 101129618, the Horizon Europe project INSTINCT under grant agreement number 101139161, and the Horizon Europe project TWIN6G under grant agreement number 101182794, as well as by the Agence Nationale de la Recherche (ANR) through the France 2030 project ANR-PEPR Networks of the Future under grant agreement NF-PERSEUS 22-PEFT-004, and by the CHIST-ERA project PASSIONATE under grant agreements CHIST-ERA-22-WAI-04 and ANR-23-CHR4-0003-01.

\bibliographystyle{IEEEtran}
\bibliography{bibliography.bib}

\begin{thebibliography}{10}
\providecommand{\url}[1]{#1}
\csname url@samestyle\endcsname
\providecommand{\newblock}{\relax}
\providecommand{\bibinfo}[2]{#2}
\providecommand{\BIBentrySTDinterwordspacing}{\spaceskip=0pt\relax}
\providecommand{\BIBentryALTinterwordstretchfactor}{4}
\providecommand{\BIBentryALTinterwordspacing}{\spaceskip=\fontdimen2\font plus
\BIBentryALTinterwordstretchfactor\fontdimen3\font minus \fontdimen4\font\relax}
\providecommand{\BIBforeignlanguage}[2]{{%
\expandafter\ifx\csname l@#1\endcsname\relax
\typeout{** WARNING: IEEEtran.bst: No hyphenation pattern has been}%
\typeout{** loaded for the language `#1'. Using the pattern for}%
\typeout{** the default language instead.}%
\else
\language=\csname l@#1\endcsname
\fi
#2}}
\providecommand{\BIBdecl}{\relax}
\BIBdecl

\bibitem{Ciuonzo2019book}
D.~Ciuonzo and P.~{Salvo Rossi}, Eds., \emph{Data Fusion in Wireless Sensor Networks: a Statistical Signal Processing Perspective}.\hskip 1em plus 0.5em minus 0.4em\relax Institution of Engineering and Technology (IET), series on Control, Robotics \& Sensors, 2019.

\bibitem{Chen2019}
Z.~Chen and Y.~Zhang, ``Providing spectrum information service using {TV} white space via distributed detection system,'' \emph{IEEE Trans. Veh. Technol.}, vol.~68, no.~8, pp. 7655--7667, 2019.

\bibitem{tabella2024bayesian}
G.~Tabella, D.~Ciuonzo, N.~Paltrinieri, and P.~Salvo~Rossi, ``Bayesian fault detection and localization through wireless sensor networks in industrial plants,'' \emph{IEEE Internet Things J.}, 2024.

\bibitem{li2007distributed}
W.~Li and H.~Dai, ``Distributed detection in wireless sensor networks using a multiple access channel,'' \emph{{IEEE} Trans. Signal Process.}, vol.~55, no.~3, pp. 822--833, 2007.

\bibitem{zhang2008optimal}
X.~Zhang, H.~V. Poor, and M.~Chiang, ``Optimal power allocation for distributed detection over {MIMO} channels in wireless sensor networks,'' \emph{{IEEE} Trans. Signal Process.}, vol.~56, no.~9, pp. 4124--4140, 2008.

\bibitem{Ciuonzo2012}
D.~Ciuonzo, G.~Romano, and {P.~Salvo Rossi}, ``Channel-aware decision fusion in distributed {MIMO} wireless sensor networks: Decode-and-fuse vs. decode-then-fuse,'' \emph{{IEEE} Trans. Wireless Commun.}, vol.~11, no.~8, pp. 2976--2985, Aug. 2012.

\bibitem{jiang2007multiuser}
M.~Jiang and L.~Hanzo, ``Multiuser {MIMO-OFDM} for next-generation wireless systems,'' \emph{Proc. IEEE}, vol.~95, no.~7, pp. 1430--1469, 2007.

\bibitem{lu2014overview}
L.~Lu, G.~Y. Li, A.~L. Swindlehurst, A.~Ashikhmin, and R.~Zhang, ``An overview of massive {MIMO}: Benefits and challenges,'' \emph{IEEE J. Sel. Topics Signal Process.}, vol.~8, no.~5, pp. 742--758, 2014.

\bibitem{ciuonzo2015}
D.~Ciuonzo, {P.~Salvo Rossi}, and S.~Dey, ``Massive {MIMO} channel-aware decision fusion,'' \emph{{IEEE} Trans. Signal Process.}, vol.~63, no.~3, pp. 604--619, 2015.

\bibitem{jiang2015massive}
F.~Jiang, J.~Chen, A.~L. Swindlehurst, and J.~A. L{\'o}pez-Salcedo, ``Massive {MIMO} for wireless sensing with a coherent multiple access channel,'' \emph{{IEEE} Trans. Signal Process.}, vol.~63, no.~12, pp. 3005--3017, 2015.

\bibitem{Chawla2021}
A.~Chawla, R.~K. Singh, A.~Patel, A.~K. Jagannatham, and L.~Hanzo, ``Distributed detection for centralized and decentralized millimeter wave massive {MIMO} sensor networks,'' \emph{IEEE Trans. Veh. Technol.}, vol.~70, no.~8, pp. 7665--7680, 2021.

\bibitem{direnzo2020}
M.~Di~Renzo, A.~Zappone, M.~Debbah, M.-S. Alouini, C.~Yuen, J.~De~Rosny, and S.~Tretyakov, ``Smart radio environments empowered by reconfigurable intelligent surfaces: How it works, state of research, and the road ahead,'' \emph{{IEEE} J. Sel. Areas Commun.}, vol.~38, no.~11, pp. 2450--2525, 2020.

\bibitem{huang2020holographic}
C.~Huang, S.~Hu, G.~C. Alexandropoulos, A.~Zappone, C.~Yuen, R.~Zhang, M.~Di~Renzo, and M.~Debbah, ``Holographic {MIMO} surfaces for {6G} wireless networks: Opportunities, challenges, and trends,'' \emph{IEEE Wireless Commun.}, vol.~27, no.~5, pp. 118--125, 2020.

\bibitem{jamali2020intelligent}
V.~Jamali, A.~M. Tulino, G.~Fischer, R.~R. M{\"u}ller, and R.~Schober, ``Intelligent surface-aided transmitter architectures for millimeter-wave ultra massive {MIMO} systems,'' \emph{IEEE Open Journal of the Communications Society}, vol.~2, pp. 144--167, 2020.

\bibitem{interdonato2024approaching}
G.~Interdonato, F.~Di~Murro, C.~D’Andrea, G.~Di~Gennaro, and S.~Buzzi, ``Approaching massive {MIMO} performance with reconfigurable intelligent surfaces: We do not need many antennas,'' \emph{IEEE Trans. Commun.}, 2024.

\bibitem{dey2020wideband}
I.~Dey, D.~Ciuonzo, and P.~Salvo~Rossi, ``Wideband collaborative spectrum sensing using massive {MIMO} decision fusion,'' \emph{{IEEE} Trans. Wireless Commun.}, vol.~19, no.~8, pp. 5246--5260, 2020.

\bibitem{dardari2020communicating}
D.~Dardari, ``Communicating with large intelligent surfaces: Fundamental limits and models,'' \emph{IEEE J. Sel. Areas Commun.}, vol.~38, no.~11, pp. 2526--2537, 2020.

\bibitem{Tenney1981}
R.~R. Tenney and N.~R. Sandell, ``Detection with distributed sensors,'' \emph{IEEE Trans. Aerosp. Electron. Syst.}, no.~4, pp. 501--510, 1981.

\bibitem{chair1986}
Z.~Chair and P.~Varshney, ``Optimal data fusion in multiple sensor detection systems,'' \emph{IEEE Trans. Aerosp. Electron. Syst.}, no.~1, pp. 98--101, 1986.

\bibitem{reibman1987}
A.~R. Reibman and L.~Nolte, ``Optimal detection and performance of distributed sensor systems,'' \emph{IEEE Trans. Aerosp. Electron. Syst.}, no.~1, pp. 24--30, 1987.

\bibitem{Chen2004}
B.~Chen, R.~Jiang, T.~Kasetkasem, and P.~K. Varshney, ``Channel aware decision fusion in wireless sensor networks,'' \emph{{IEEE} Trans. Signal Process.}, vol.~52, no.~12, pp. 3454--3458, Dec. 2004.

\bibitem{Mergen2007}
G.~Mergen, V.~Naware, and L.~Tong, ``Asymptotic detection performance of type-based multiple access over multiaccess fading channels,'' \emph{{IEEE} Trans. Signal Process.}, vol.~55, no.~3, pp. 1081--1092, 2007.

\bibitem{Yiu2008}
S.~Yiu and R.~Schober, ``Nonorthogonal transmission and noncoherent fusion of censored decisions,'' \emph{IEEE Trans. Veh. Technol.}, vol.~58, no.~1, pp. 263--273, 2008.

\bibitem{zhang2008}
X.~Zhang, H.~V. Poor, and M.~Chiang, ``Optimal power allocation for distributed detection over {MIMO} channels in wireless sensor networks,'' \emph{{IEEE} Trans. Signal Process.}, vol.~56, no.~9, pp. 4124--4140, 2008.

\bibitem{Appadwedula2005}
S.~Appadwedula, V.~V. Veeravalli, and D.~L. Jones, ``Energy-efficient detection in sensor networks,'' \emph{IEEE J. Sel. Areas Commun.}, vol.~23, no.~4, pp. 693--702, 2005.

\bibitem{ahmadi2009}
H.~R. Ahmadi and A.~Vosoughi, ``Channel aware sensor selection in distributed detection systems,'' in \emph{IEEE 10th Workshop on Signal Processing Advances in Wireless Communications (SPAWC)}, 2009.

\bibitem{Ciuonzo2019pimrc}
D.~Ciuonzo, G.~Gelli, A.~Pescap{\'e}, and F.~Verde, ``Decision fusion rules in ambient backscatter wireless sensor networks,'' in \emph{IEEE 30th Annual Int. Symp. on Personal, Indoor and Mobile Radio Communications (PIMRC)}, 2019.

\bibitem{Tarighati2017}
A.~Tarighati, J.~Gross, and J.~Jald{\'e}n, ``Decentralized hypothesis testing in energy harvesting wireless sensor networks,'' \emph{{IEEE} Trans. Signal Process.}, vol.~65, no.~18, pp. 4862--4873, 2017.

\bibitem{chawla2019}
A.~Chawla, A.~Patel, A.~K. Jagannatham, and P.~K. Varshney, ``Distributed detection in massive {MIMO} wireless sensor networks under perfect and imperfect {CSI},'' \emph{{IEEE} Trans. Signal Process.}, vol.~67, no.~15, pp. 4055--4068, 2019.

\bibitem{zappone2022surface}
A.~Zappone, M.~Di~Renzo, and R.~K. Fotock, ``Surface-based techniques for {IoT} networks: Opportunities and challenges,'' \emph{IEEE Internet of Things Magazine}, vol.~5, no.~4, pp. 72--77, 2022.

\bibitem{Sahin23}
A.~Şahin and R.~Yang, ``A survey on over-the-air computation,'' \emph{IEEE Commun. Surveys Tuts}, vol.~25, no.~3, pp. 1877--1908, 2023.

\bibitem{fang2021}
W.~Fang, Y.~Jiang, Y.~Shi, Y.~Zhou, W.~Chen, and K.~B. Letaief, ``Over-the-air computation via reconfigurable intelligent surface,'' \emph{IEEE Trans. Commun.}, vol.~69, no.~12, pp. 8612--8626, 2021.

\bibitem{zhang2022worst}
W.~Zhang, J.~Xu, W.~Xu, X.~You, and W.~Fu, ``Worst-case design for {RIS-aided} over-the-air computation with imperfect {CSI},'' \emph{{IEEE} Commun. Lett.}, vol.~26, no.~9, pp. 2136--2140, 2022.

\bibitem{zhai2022beamforming}
X.~Zhai, G.~Han, Y.~Cai, and L.~Hanzo, ``Beamforming design based on two-stage stochastic optimization for {RIS-assisted} over-the-air computation systems,'' \emph{IEEE Internet Things J.}, vol.~9, no.~7, pp. 5474--5488, 2022.

\bibitem{zhao2023ris}
Y.~Zhao, W.~Xu, and X.~Ye, ``{RIS-Assisted} {CSIT-free} data fusion with timing misalignment,'' in \emph{9th IEEE International Conference on Computer and Communication Engineering (ICCCE)}, 2023, pp. 12--17.

\bibitem{zhang2023beamforming}
D.~Zhang, M.~Xiao, M.~Skoglund, and H.~V. Poor, ``Beamforming design for active {RIS-Aided} over-the-air computation,'' \emph{arXiv preprint arXiv:2311.18418}, 2023.

\bibitem{li2021double}
J.~Li, M.~Fu, Y.~Zhou, and Y.~Shi, ``{Double-RIS} assisted over-the-air computation,'' in \emph{IEEE Globecom Workshops (GC Wkshps)}, 2021, pp. 1--6.

\bibitem{zhai2022joint}
X.~Zhai, G.~Han, Y.~Cai, and L.~Hanzo, ``Joint beamforming aided over-the-air computation systems relying on both {BS-side} and user-side reconfigurable intelligent surfaces,'' \emph{{IEEE} Trans. Wireless Commun.}, vol.~21, no.~12, pp. 10\,766--10\,779, 2022.

\bibitem{zhai2023simultaneously}
X.~Zhai, G.~Han, Y.~Cai, Y.~Liu, and L.~Hanzo, ``Simultaneously transmitting and reflecting {(STAR) RIS} assisted over-the-air computation systems,'' \emph{IEEE Trans. Commun.}, vol.~71, no.~3, pp. 1309--1322, 2023.

\bibitem{hu2022ris}
L.~Hu, Z.~Wang, H.~Zhu, Y.~Shi, and Y.~Zhou, ``{RIS-assisted} over-the-air computation in millimeter wave communication networks,'' in \emph{IEEE 95th Vehicular Technology Conference (VTC2022-Spring)}, 2022, pp. 1--5.

\bibitem{mao2022intelligent}
S.~Mao, N.~Zhang, J.~Hu, and K.~Yang, ``Intelligent reflecting surface-assisted over-the-air computation for backscatter sensor networks,'' \emph{IEEE Trans. Veh. Technol.}, vol.~72, no.~5, pp. 6839--6843, 2023.

\bibitem{wang2021wireless}
Z.~Wang, Y.~Shi, Y.~Zhou, H.~Zhou, and N.~Zhang, ``Wireless-powered over-the-air computation in intelligent reflecting surface-aided {IoT} networks,'' \emph{IEEE Internet Things J.}, vol.~8, no.~3, pp. 1585--1598, 2021.

\bibitem{mao2024joint}
S.~Mao, N.~Zhang, L.~Liu, T.~Liu, J.~Hu, K.~Yang, and D.~Niyato, ``Joint beamforming and reflecting design for {IRS-Aided} wireless powered over-the-air computation and communication networks,'' \emph{IEEE Trans. Commun.}, vol.~72, no.~4, pp. 2216--2231, 2024.

\bibitem{Ahmed2022}
M.~F. Ahmed, K.~P. Rajput, N.~K. Venkategowda, K.~V. Mishra, and A.~K. Jagannatham, ``Joint transmit and reflective beamformer design for secure estimation in {IRS-aided WSNs},'' \emph{IEEE Signal Process. Lett.}, vol.~29, pp. 692--696, 2022.

\bibitem{ge2024ris}
J.~Ge, Y.-C. Liang, S.~Wang, and C.~Sun, ``{RIS-Assisted} cooperative spectrum sensing for cognitive radio networks,'' \emph{{IEEE} Trans. Wireless Commun.}, vol.~23, no.~9, pp. 12\,547--12\,562, 2024.

\bibitem{rajput2024joint}
K.~P. Rajput, L.~Wu, M.~B. Shankar, and P.~K. Varshney, ``Joint transmit precoders and passive reflection beamformer design in {IRS-Aided IoT} networks,'' in \emph{IEEE International Conference on Acoustics, Speech and Signal Processing (ICASSP)}, 2024, pp. 156--160.

\bibitem{mudkey2022wireless}
N.~Mudkey, D.~Ciuonzo, A.~Zappone, and P.~Salvo~Rossi, ``Wireless inference gets smarter: {RIS-}assisted channel-aware {MIMO} decision fusion,'' in \emph{IEEE 12th Sensor Array and Multichannel Signal Processing Workshop (SAM)}, 2022, pp. 26--30.

\bibitem{Ciuonzo2025icassp}
D.~Ciuonzo, A.~Zappone, M.~Di~Renzo, and L.~Wu, ``Massive {MIMO} channel-aware decision fusion aided by reconfigurable intelligent surfaces,'' in \emph{IEEE International Conference on Acoustics, Speech, and Signal Processing (ICASSP)}, 2025.

\bibitem{sun2016majorization}
Y.~Sun, P.~Babu, and D.~P. Palomar, ``Majorization-minimization algorithms in signal processing, communications, and machine learning,'' \emph{{IEEE} Trans. Signal Process.}, vol.~65, no.~3, pp. 794--816, 2016.

\bibitem{salvorossi2013}
{P.~Salvo Rossi}, D.~Ciuonzo, and G.~Romano, ``Orthogonality and cooperation in collaborative spectrum sensing through {MIMO} decision fusion,'' \emph{{IEEE} Trans. Wireless Commun.}, vol.~12, no.~11, pp. 5826--5836, Nov. 2013.

\bibitem{albreem2017green}
M.~A. Albreem, A.~A. El-Saleh, M.~Isa, W.~Salah, M.~Jusoh, M.~Azizan, and A.~Ali, ``Green {Internet of Things (IoT)}: An overview,'' in \emph{IEEE 4th International Conference on Smart Instrumentation, Measurement and Application (ICSIMA)}, 2017, pp. 1--6.

\bibitem{aldalahmeh2022}
S.~A. Aldalahmeh and D.~Ciuonzo, ``Distributed detection fusion in clustered sensor networks over multiple access fading channels,'' \emph{IEEE Transactions on Signal and Information Processing over Networks}, vol.~8, pp. 317--329, 2022.

\bibitem{ellingson2021path}
S.~W. Ellingson, ``Path loss in reconfigurable intelligent surface-enabled channels,'' in \emph{IEEE 32nd Annual International Symposium on Personal, Indoor and Mobile Radio Communications (PIMRC)}, 2021, pp. 829--835.

\bibitem{tang2020wireless}
W.~Tang, M.~Z. Chen, X.~Chen, J.~Y. Dai, Y.~Han, M.~Di~Renzo, Y.~Zeng, S.~Jin, Q.~Cheng, and T.~J. Cui, ``Wireless communications with reconfigurable intelligent surface: Path loss modeling and experimental measurement,'' \emph{{IEEE} Trans. Wireless Commun.}, vol.~20, no.~1, pp. 421--439, 2020.

\bibitem{tang2022path}
W.~Tang, X.~Chen, M.~Z. Chen, J.~Y. Dai, Y.~Han, M.~Di~Renzo, S.~Jin, Q.~Cheng, and T.~J. Cui, ``Path loss modeling and measurements for reconfigurable intelligent surfaces in the millimeter-wave frequency band,'' \emph{IEEE Trans. Commun.}, vol.~70, no.~9, pp. 6259--6276, 2022.

\bibitem{abrardo2021intelligent}
A.~Abrardo, D.~Dardari, and M.~Di~Renzo, ``Intelligent reflecting surfaces: Sum-rate optimization based on statistical position information,'' \emph{IEEE Trans. Commun.}, vol.~69, no.~10, pp. 7121--7136, 2021.

\bibitem{DegliEsposti2022}
V.~Degli-Esposti, E.~M. Vitucci, M.~D. Renzo, and S.~A. Tretyakov, ``Reradiation and scattering from a reconfigurable intelligent surface: A general macroscopic model,'' \emph{IEEE Trans. Antennas Propag.}, vol.~70, no.~10, pp. 8691--8706, 2022.

\bibitem{feng2023near}
C.~Feng, H.~Lu, Y.~Zeng, T.~Li, S.~Jin, and R.~Zhang, ``Near-field modelling and performance analysis for extremely large-scale {IRS} communications,'' \emph{{IEEE} Trans. Wireless Commun.}, 2023.

\bibitem{Kay1998}
S.~M. Kay, ``Fundamentals of statistical signal processing, vol. ii: Detection theory,'' \emph{Signal Processing. Upper Saddle River, NJ: Prentice Hall}, 1998.

\bibitem{Picinbono1995}
B.~Picinbono, ``On deflection as a performance criterion in detection,'' \emph{IEEE Trans. Aerosp. Electron. Syst.}, vol.~31, no.~3, pp. 1072--1081, 1995.

\bibitem{Quan2008}
Z.~Quan, S.~Cui, and A.~H. Sayed, ``Optimal linear cooperation for spectrum sensing in cognitive radio networks,'' \emph{{IEEE} J. Sel. Topics Signal Process.}, vol.~2, no.~1, pp. 28--40, Feb. 2008.

\bibitem{lei2010coherent}
A.~Lei and R.~Schober, ``Coherent max-log decision fusion in wireless sensor networks,'' \emph{{IEEE} Trans. Commun.}, vol.~58, no.~5, pp. 1327--1332, 2010.

\bibitem{zappone2022energy}
A.~Zappone, B.~Matthiesen, and A.~Dekorsy, ``Energy efficiency of holographic transceivers based on {RIS},'' in \emph{IEEE Global Communications Conference (GLOBECOM)}, 2022, pp. 4613--4618.

\end{thebibliography}

\begin{IEEEbiography}[{\includegraphics[width=1in,height=1.25in,clip,keepaspectratio]{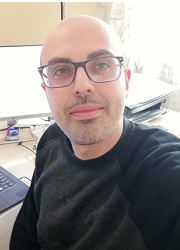}}]{Domenico~Ciuonzo} (Senior Member, IEEE)  is a Tenure-track Assistant Professor at University of Naples \say{Federico II}. 
He holds a Ph.D. degree from the University of Campania \say{L. Vanvitelli}, Italy. 
Since 2011, he has been holding several visiting researcher appointments (NATO CMRE, UConn, NTNU, CTTC). 
He is the recipient of two Best Paper awards (IEEE ICCCS 2019 and Elsevier Computer Networks 2020), the 2019 Exceptional Service award from \textsc{IEEE AESS}, the 2020 Early-Career Technical Achievement award from \textsc{IEEE Sensors Council} for sensor networks/systems and the 2021 Early-Career Award from \textsc{IEEE AESS} for contributions to decentralized inference and sensor fusion in networked sensor systems.
Domenico serves as editor of the  \textsc{IEEE Internet of Things Journal}, lead editor for \textsc{IEEE Internet of Things Magazine}, and has served as executive editor for the \textsc{IEEE Communications Letters} and area editor for  the \textsc{IEEE Transactions on Aerospace and Electronic Systems}.
His research interests include data fusion, statistical signal processing, wireless sensor networks, the IoT, and machine learning.
\end{IEEEbiography}

\begin{IEEEbiography}[{\includegraphics[width=1in,height=1.25in,clip,keepaspectratio]{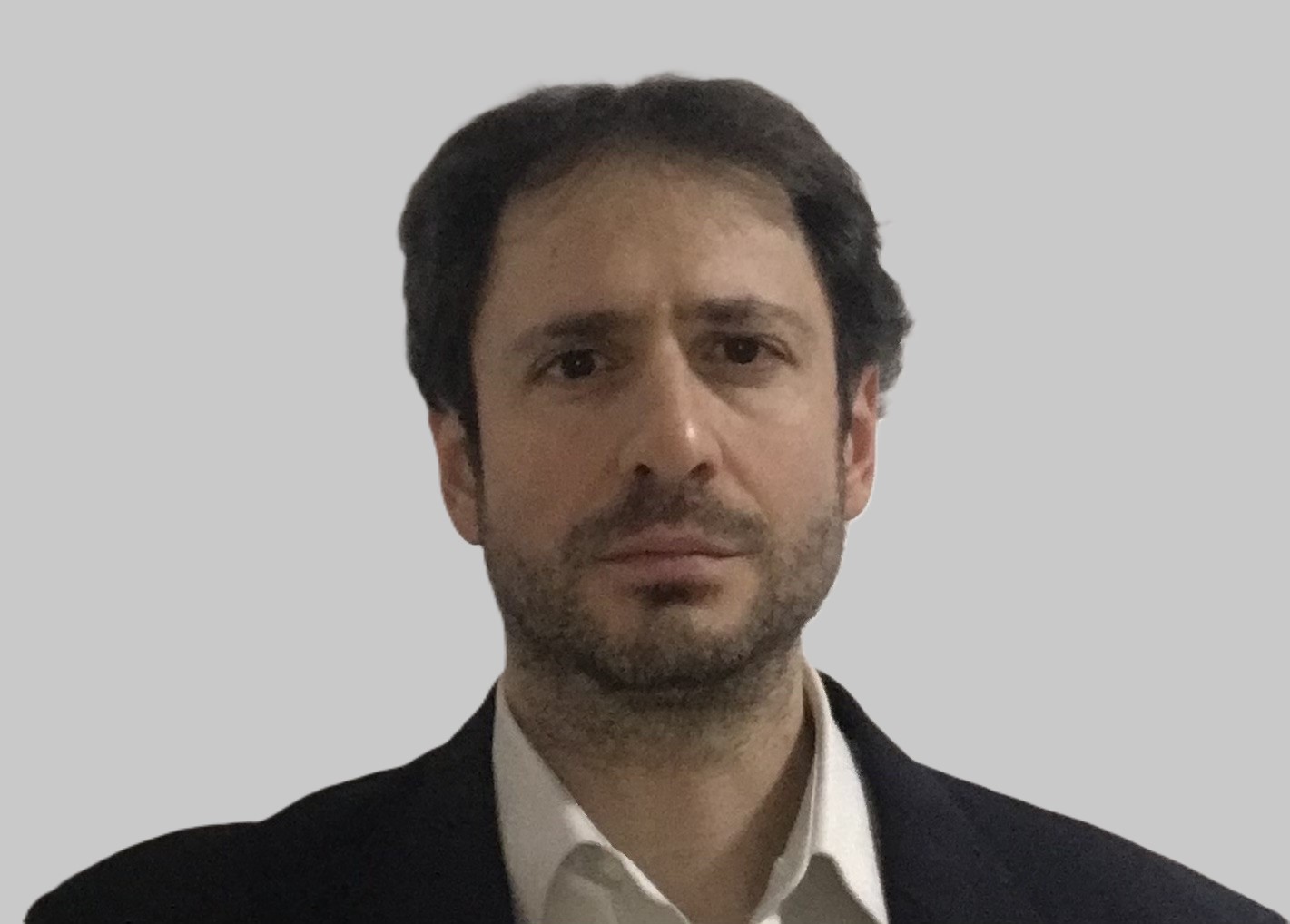}}]{Alessio Zappone} (Fellow, IEEE) is a tenured professor at the University of Cassino and Southern Lazio. His research interests lie in the area of communication theory and signal processing, with main focus on optimization techniques for resource allocation and energy efficiency maximization. For his research, Alessio received several award among which the IEEE Marconi Prize Paper Award in Wireless Communications in 2021, the IEEE Communications Society Fred W. Ellersick Prize in 2023, and the IEEE Communications Society Best Tutorial Paper Award in 2024. Alessio serves as editor of the \textsc{IEEE Transactions on Wireless Communications}, area editor of the \textsc{IEEE Communications Letters}, and has served as senior editor of the \textsc{IEEE Signal Processing Letters} and guest editor of two \textsc{IEEE Journal on Selected Areas in Communications} special issues.
\end{IEEEbiography}

\begin{IEEEbiography}[{\includegraphics[width=1in,height=1.25in,clip,keepaspectratio]{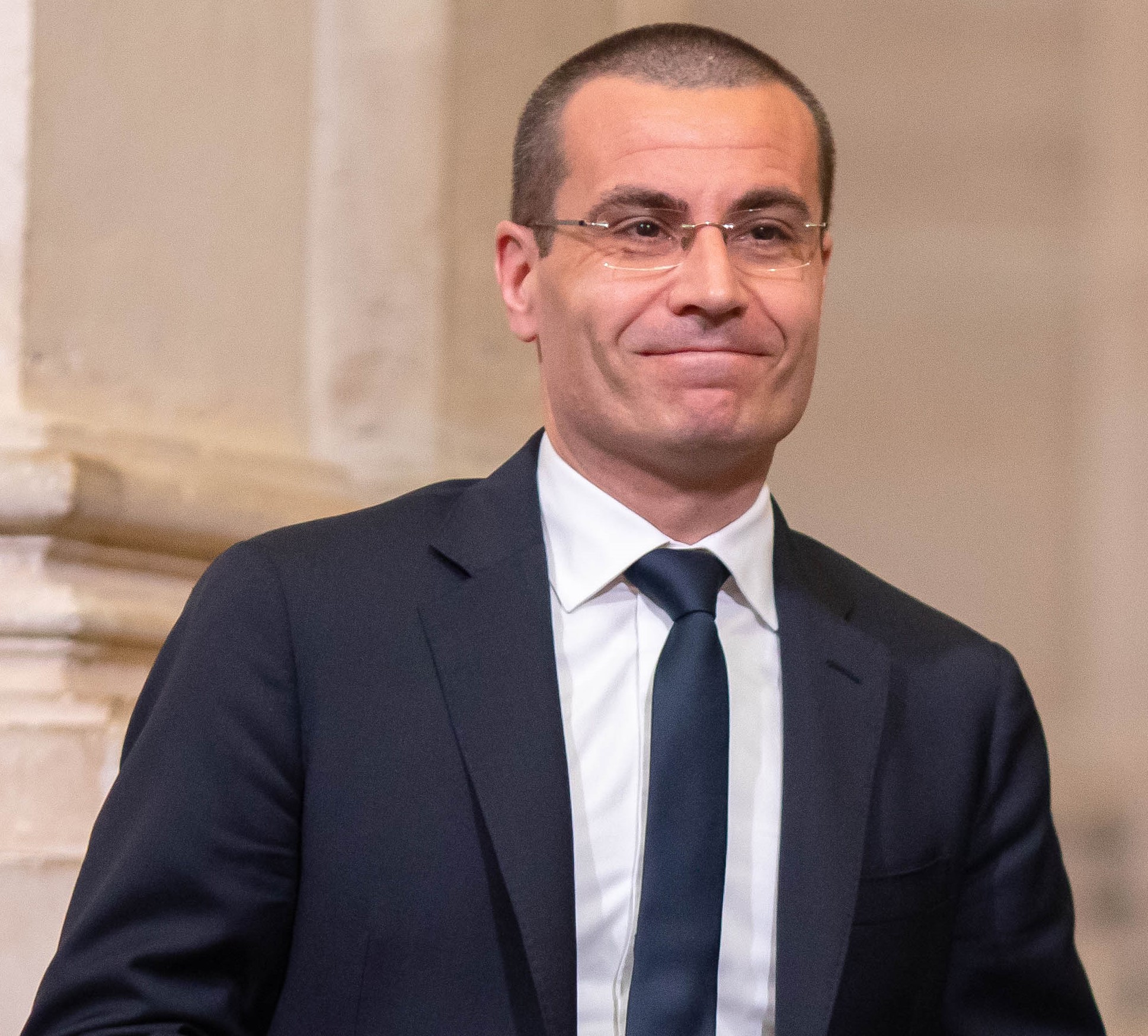}}]{Marco~Di~Renzo} (Fellow, IEEE) received the Laurea (cum laude) and Ph.D. degrees in electrical engineering from the University of L’Aquila, Italy, in 2003 and 2007, respectively, and the Habilitation à Diriger des Recherches (Doctor of Science) degree from University Paris-Sud (currently Paris-Saclay University), France, in 2013. Currently, he is a CNRS Research Director (Professor) and the Head of the Intelligent Physical Communications group with the Laboratory of Signals and Systems (L2S) at CNRS \& CentraleSupélec, Paris-Saclay University, Paris, France, as well as a Chair Professor in Telecommunications Engineering with the Centre for Telecommunications Research -- Department of Engineering, King’s College London, London, United Kingdom. He was a France-Nokia Chair of Excellence in ICT at the University of Oulu (Finland), a Tan Chin Tuan Exchange Fellow in Engineering at Nanyang Technological University (Singapore), a Fulbright Fellow at The City University of New York (USA), a Nokia Foundation Visiting Professor at Aalto University (Finland), and a Royal Academy of Engineering Distinguished Visiting Fellow at Queen’s University Belfast (U.K.). He is a Fellow of the IEEE, IET, EURASIP, and AAIA; an Academician of AIIA; an Ordinary Member of the European Academy of Sciences and Arts, an Ordinary Member of the Academia Europaea; an Ambassador of the European Association on Antennas and Propagation; and a Highly Cited Researcher. His recent research awards include the Michel Monpetit Prize conferred by the French Academy of Sciences, the IEEE Communications Society Heinrich Hertz Award, and the IEEE Communications Society Marconi Prize Paper Award in Wireless Communications. He served as the Editor-in-Chief of IEEE Communications Letters from 2019 to 2023. His current main roles within the IEEE Communications Society include serving as a Voting Member of the Fellow Evaluation Standing Committee, as the Chair of the Publications Misconduct Ad Hoc Committee, and as the Director of Journals. Also, he is on the Editorial Board of the Proceedings of the IEEE.
\end{IEEEbiography}

\end{document}